\documentclass{article}
\usepackage[T1]{fontenc}
\usepackage{iclr2027_conference,times}
\usepackage{amsmath,graphicx,hyperref,url}
\hypersetup{hidelinks}
\usepackage{booktabs}
\usepackage{tabularx}
\iclrfinalcopy   
\title{SURE-Tagger: Structured Speech Representations for Compositional Evaluation and Targeted Data Selection}
\author{%
\parbox[t]{\dimexpr\textwidth-2\tabcolsep\relax}{%
\centering
\bfseries
Jing Peng\textsuperscript{1,*}\quad
Huifei Wang\textsuperscript{2,3,*}\quad
Weihan Chen\textsuperscript{1,*}\quad
Shuo Gao\textsuperscript{1}\quad
Bicheng Zhang\textsuperscript{1}\\[0.4em]
Junhao Du\textsuperscript{1}\quad
Hanqi Li\textsuperscript{1}\quad
Chaolei Liu\textsuperscript{1}\quad
Wen Wen\textsuperscript{1}\quad
Tao Liu\textsuperscript{2}\\[0.4em]
Xu Li\textsuperscript{4}\quad
Yu Xi\textsuperscript{1}\quad
Shuai Wang\textsuperscript{5}\quad
Kai Yu\textsuperscript{1,$\dagger$}\\[0.9em]
\normalfont\small
\textsuperscript{1}X-LANCE Lab, Department of Computer Science and Engineering,
Shanghai Jiao Tong University,\\
MoE Key Lab of Artificial Intelligence, Jiangsu Key Lab of Language Computing\\[0.35em]
\textsuperscript{2}AISpeech Ltd.\quad
\textsuperscript{3}Shenzhen University
\textsuperscript{4}Independent Researcher\quad
\textsuperscript{5}Nanjing University\\
\footnotesize
\textsuperscript{*}Equal contribution.\quad
\textsuperscript{$\dagger$}Corresponding author
}%
}

\begin{document}
%
\maketitle
\lhead{Preprint}
\begin{abstract}
Speech conveys linguistic content, speaker characteristics, temporal
interactions, paralinguistic cues, and acoustic environments.
While recent speech captioning models describe these dimensions in
natural language, inconsistent attribute coverage and ambiguous omissions
hinder systematic analysis across recordings.
We introduce \textbf{SURE-Tagger}, a framework that constructs
semi-structured speech representations with explicit attribute values,
supplementary descriptions, and abstention under insufficient evidence.
By integrating specialist observations through language-model reasoning,
SURE-Tagger enables speech conditions to be consistently identified
and jointly queried across datasets.
We further introduce \textbf{Tag-Bench}, combining real recordings
with controlled, agentically synthesized speech scenes and verified
attribute annotations.
On Tag-Bench, SURE-Tagger achieves the highest scores on
\textbf{25 of 29} evaluated dimensions for real speech and
\textbf{24 of 29} for synthetic speech, outperforming caption-based
baselines across structural and descriptive attributes.
Beyond annotation, we use these representations to construct
cross-dataset evaluation subsets and reveal condition-specific
ASR errors.
Using a separate candidate pool, SURE-Tagger-guided
selection reduces the overall ASR error rate of
Qwen3-ASR-1.7B by 14.3\% relative to the unadapted
model under a one-hour fine-tuning budget, while
caption-guided and random selection increase error
in the same setting.
These results demonstrate how structured speech representations
can connect rich annotation, compositional evaluation, and
targeted model improvement.
\end{abstract}
\par

\section{Introduction}
\label{sec:introduction}

\begin{figure}[t]
    \centering
    \includegraphics[width=0.6\linewidth]{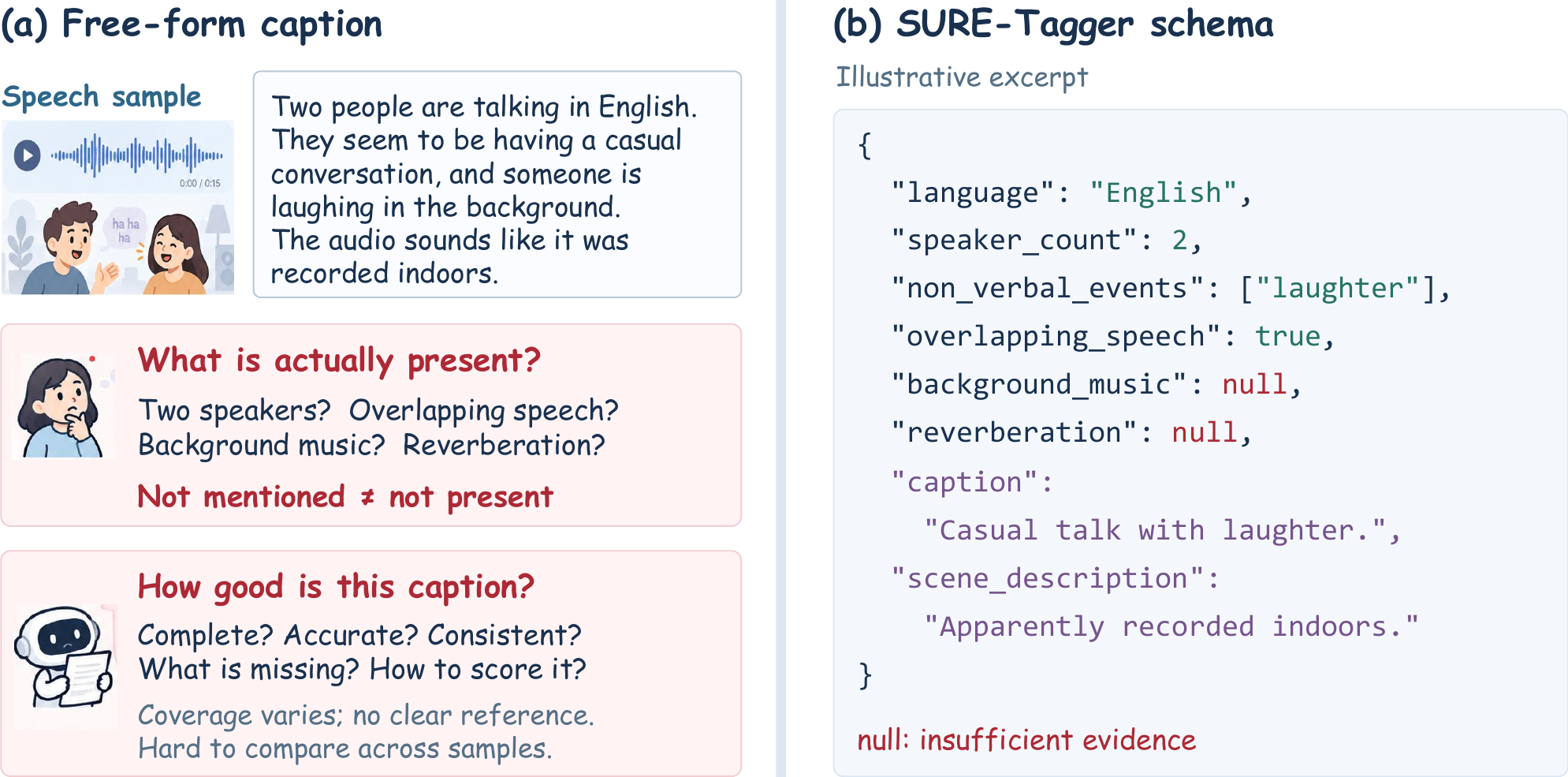}
    \caption{
        Free-form speech captions and a semi-structured
        speech representation.
        Left: ambiguities in attribute coverage and omission.
        Right: an illustrative schema excerpt with explicit
        attribute values and supplementary descriptions;
        \texttt{null} denotes insufficient evidence.
    }
    \label{fig:motivation}
\end{figure}

Speech conveys linguistic content, speaker characteristics
and interactions, paralinguistic cues, and acoustic conditions~\citep{peng2025survey}.
Recent speech and audio captioning models describe these
dimensions in natural language, capturing information beyond
conventional transcription~\citep{xu2023secap,jin2024speechcraft,ma2026omnicaptioner}.
However, free-form descriptions do not consistently specify
the same attributes across recordings: their coverage and
wording vary, and an omitted attribute may be absent,
overlooked, or uncertain.
As illustrated in Figure~\ref{fig:motivation}(a),
such omissions cannot be interpreted as confirmed absence,
making captions difficult to use directly as consistent,
sample-level attribute records.

This limitation becomes particularly consequential when
evaluating speech models beyond aggregate dataset scores.
Recordings from different datasets may share speech
conditions under which model performance differs, yet
dataset-level results can obscure these differences,
especially when multiple conditions co-occur
~\citep{sagawa2020groupdro,oakdenrayner2020hidden}.
We therefore study \emph{compositional evaluation}:
constructing evaluation subsets through conjunctions of
speech attributes across datasets and measuring model
performance on the resulting subsets.
For example, identifying recordings with multiple speakers
and overlapping speech requires both attributes to be
explicitly established, as illustrated in
Figure~\ref{fig:motivation}(b).
This motivates a shared representation that distinguishes
confirmed attribute values, including absence where
applicable, from insufficient evidence, while allowing
attributes to be jointly queried across recordings.
The resulting condition-wise error profiles can further
guide targeted data selection for model adaptation.

Existing work provides complementary foundations for this goal.
Omni-Captioner advances detailed audio and audio-visual
description, while Omni-Cloze assesses the information conveyed
by such descriptions through fine-grained cloze
questions~\citep{ma2026omnicaptioner}.
However, detailed captions do not necessarily specify explicit
values for every attribute required by a downstream query.
Structured metadata offers another route:
AudioCards organizes acoustic attributes and sonic descriptors
for sound design~\citep{sridhar2026audiocards}.
Vox-Profile characterizes static speaker traits and dynamic
speech properties, including their use in analyzing ASR
performance variability~\citep{feng2025voxprofile}.
Dynamic-SUPERB provides broad coverage of speech and audio
capabilities through a diverse collection of evaluation
tasks~\citep{huang2025dynamicsuperb}.
Complementing task coverage, SURE standardizes prediction formats,
normalization, and scoring for reproducible speech-understanding
experiments~\citep{peng2026unified}.
Building on these efforts, we focus on connecting rich speech
annotation with compositional evaluation and targeted data
selection through a shared, sample-level representation.

To this end, we introduce SURE-Tagger, a framework
for constructing semi-structured, sample-level speech
representations. \emph{Each representation combines explicit
values for predefined speech attributes with open-ended
descriptions that preserve information beyond their
value spaces.} Unlike free-form captions, the explicit
attributes distinguish confirmed values, including
absence where applicable, from insufficient evidence
represented by \texttt{null}. This design retains
descriptive flexibility while making speech conditions
explicit and computationally accessible. To construct the representation, specialized perception
models provide attribute-specific observations, which
SURE-Tagger treats as evidence rather than direct schema
assignments. A language-model agent integrates these
observations under the schema's field definitions and
constraints, reconciling conflicting indications where
possible and retaining \texttt{null} when evidence
remains insufficient. The resulting representations
enable consistent comparison and joint querying of
speech conditions across recordings and datasets.

We evaluate both the reliability and downstream utility
of these representations.
We introduce Tag-Bench, combining real recordings with
controlled, agentically synthesized speech scenes and
verified attribute annotations to assess annotation quality.
We further examine attribute preservation through speech
reconstruction.
Using recordings from ten heterogeneous datasets, we
construct schema-based evaluation subsets and analyze
condition-wise errors across multiple ASR models.
These error profiles then guide training-data selection
from a separate candidate pool.
With just one hour of fine-tuning data, SURE-Tagger-guided
selection reduces the held-out WER of Qwen3-ASR-1.7B
by 14.3\% relative to the unadapted model, while caption-guided selection and 
random sampling instead increase WER. Our main contributions are as follows:
\begin{enumerate}
   \item \textbf{Semi-structured speech representation.}
    We introduce a sample-level representation that combines
    explicit, jointly queryable speech attributes, abstention
    states, and supplementary open-ended descriptions.
    We instantiate this representation through SURE-Tagger,
    an annotation framework that constructs the schema from specialized speech observations.

    \item \textbf{Tag-Bench: controlled evaluation of rich speech
    annotation.}
    We introduce Tag-Bench, combining real recordings with
    agentically synthesized, constraint-validated speech scenes
    to evaluate annotation accuracy and coverage.

    \item \textbf{Compositional evaluation and diagnosis-guided
    data selection.}
    We demonstrate how SURE-Tagger representations support
    cross-dataset evaluation under joint speech conditions
    and enable condition-wise error analysis to guide targeted
    data selection for ASR adaptation.
\end{enumerate}

\section{SURE-Tagger}
\label{sec:sure_tagger}

\subsection{Semi-structured Speech Representation}
As illustrated in Figure~\ref{fig:schema_state},
the representation organizes speech information into
five semantic layers: lexical content, speaker
characteristics, temporal interactions, paralinguistic
and non-verbal information, and acoustic scenes and events.
Each layer contains attributes with defined semantics
and value spaces, while supplementary descriptions
retain details not fully captured by these fields.
The layers provide complementary views of the same
recording rather than independent scene labels:
for example, speaker count and speech overlap describe
different but jointly relevant aspects of a multi-speaker
interaction. Their explicit values can therefore be
combined to define evaluation conditions across recordings.
The schema defines a set of explicit attributes
$\mathcal{K}=\{k_1,\ldots,k_K\}$, each with a predefined
value space $\mathcal{V}_i$.
To distinguish insufficient evidence from confirmed attribute
states, each attribute also admits a \texttt{null} value:
\begin{equation}
v_i(x)\in\mathcal{V}_i\cup\{\texttt{null}\}.
\end{equation}
Here, \texttt{null} indicates that the available evidence does
not support a reliable assignment, whereas confirmed presence
or absence is represented by the corresponding value in
$\mathcal{V}_i$, where applicable.

The schema also retains supplementary descriptions for
information that cannot be fully captured by predefined
attribute values.
For an audio recording $x$, the final representation is
defined as
\begin{equation}
S^\ast(x)=
\left(
\{(k_i,v_i(x))\}_{i=1}^{K},
d(x)
\right),
\end{equation}
where $d(x)$ denotes the supplementary descriptions.
Explicit attributes provide consistent values for
compositional queries, while descriptive fields preserve
open-ended information beyond the predefined value spaces.

\subsection{Evidence-Grounded Construction}
\label{sec:evidence_construction}

To instantiate the proposed representation, SURE-Tagger
collects attribute-specific observations from specialized
perception models and uses an LLM to construct the
final schema under its field definitions and constraints.
This design separates the acquisition of speech evidence
from its assignment to the shared representation.

For an input recording $x$, the perception models produce
a collection of observations
$\mathcal O(x)=\{o_1(x),\ldots,o_M(x)\}$,
each retaining its source information.
The models may be executed conditionally: FireRedLID~\citep{xu2026firered}
provides an initial language prediction that determines
the diarization model, with MOSS-Diarize~\citep{yu2026moss} used for
Chinese and English and DiariZen~\citep{han2025diarizen} for other languages.
FireRedAED and ASR provide additional observations
of acoustic events and linguistic content, respectively.

Before final schema construction, we apply a language
consistency check using the ASR transcript.
If more than 90\% of the transcribed words belong
to a single language different from the initial
prediction, the language value is revised when both
languages belong to the same diarization-model group.
Otherwise, when the conflicting languages require
different groups, the value is set to \texttt{null}.
The original observations and the check results
are retained for subsequent schema construction.

The LLM then constructs the final representation:
\begin{equation}
S^*(x)=\Phi\bigl(\mathcal O(x),\mathcal C\bigr),
\label{eq:schema_construction}
\end{equation}
where $\mathcal C$ denotes the schema definitions
and constraints, and $\mathcal O(x)$ includes the
collected observations and consistency-check results.
The LLM uses this evidence to assign explicit
attribute values and generate supplementary
descriptions, retaining \texttt{null} when the
available evidence is insufficient.
The final schema is produced in a single LLM call.

\subsection{Compositional Evaluation and Targeted Data Selection}

\begin{figure}[t]
    \centering
    \includegraphics[width=\textwidth]{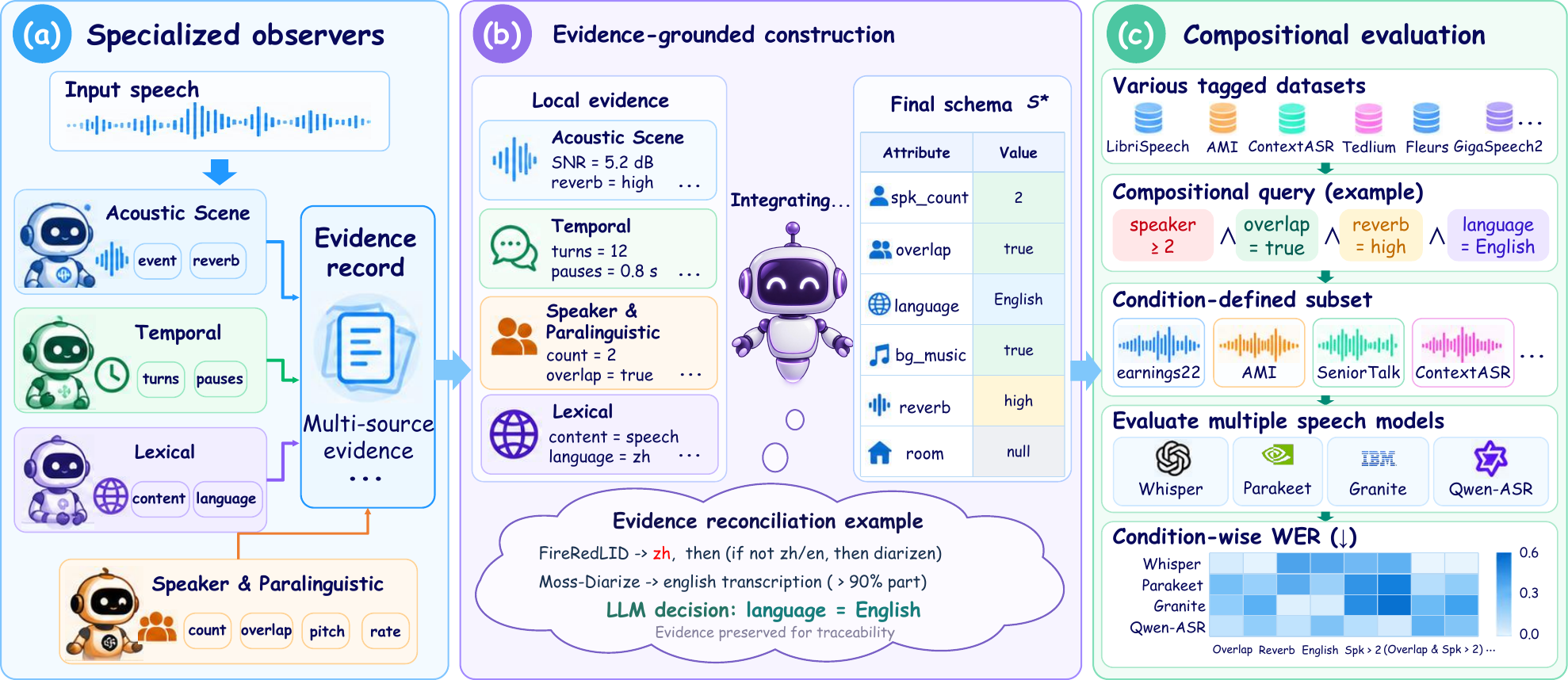}
    \caption{
    Overview of SURE-Tagger.
    The semi-structured representation combines
    explicit speech attributes with supplementary descriptions.
    SURE-Tagger constructs it by integrating observations
    from specialized perception models, enabling
    attribute-based analysis across recordings.
    }
    \label{fig:schema_state}
\end{figure}

As illustrated in Figure~\ref{fig:schema_state}(c),
the resulting schemas enable recordings from heterogeneous
datasets to be reorganized into evaluation subsets defined
by shared speech conditions.
A compositional query $q$ specifies predicates over selected
attributes:
\begin{equation}
q=\{(k_i,c_i)\}_{i\in I},
\end{equation}
where $c_i$ denotes a predicate over the value of attribute $k_i$.
The corresponding evaluation subset is defined as
\begin{equation}
\mathcal D_q
=
\{x\in\mathcal D\mid S^\ast(x)\models q\},
\end{equation}
where $\mathcal D$ denotes the collection of recordings
across datasets.
A recording is included only when all specified conditions
are confirmed by its explicit attribute values.
Attributes assigned \texttt{null} do not satisfy these
conditions and are not treated as confirmed negative cases.

For example, the query in Figure~\ref{fig:schema_state}(c)
selects English recordings with at least two speakers,
overlapping speech, and high reverberation.
For a speech model $m$, its performance under condition $q$
is measured as
\begin{equation}
P(m,q)=\operatorname{Metric}(m,\mathcal D_q),
\end{equation}
where $\operatorname{Metric}$ is a task-specific evaluation
metric, such as corpus-level WER for ASR.
Comparing performance across individual and composed
conditions provides a condition-wise view of model errors
beyond dataset-level aggregate scores.

Beyond evaluation, the same representations support
targeted data selection.
We cluster recordings in a diagnosis set according to their
schema attributes and estimate the target model's WER
within each cluster.
These error estimates guide sampling from a separate
training candidate pool represented using the same schema,
with higher sampling weights assigned to speech conditions
associated with greater recognition errors.
This connects condition-wise model diagnosis with
training-data selection under a fixed adaptation budget.

\section{Tag-Bench}
\label{sec:tag_bench}

We introduce Tag-Bench to evaluate rich speech annotation
under individual and composed attribute conditions.
It combines 500 controlled synthetic scenes with 650 recordings
from existing corpora, including AMI~\citep{carletta2006ami},
FLEURS~\citep{conneau2022fleurs}, LibriSpeech~\citep{panayotov2015librispeech},
DESED~\citep{turpault2019desed}, and FSD50K~\citep{fonseca2020fsd50k}.
The existing recordings provide diverse speech and audio
conditions, while synthetic scenes enable targeted coverage
of attribute combinations that may be underrepresented
in these corpora.

\begin{figure}[t]
    \centering
    \includegraphics[width=\textwidth]{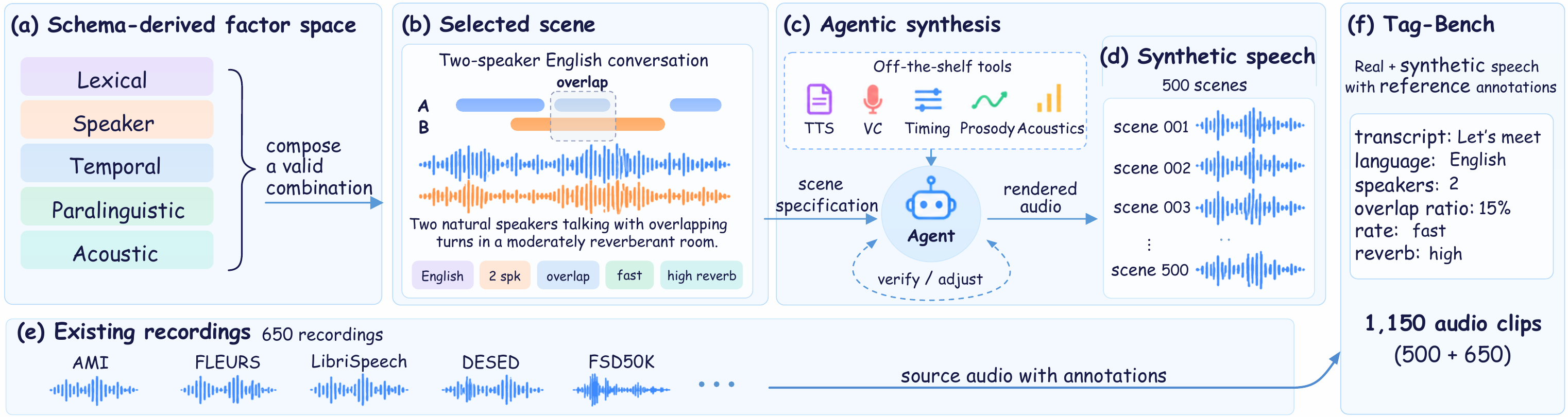}
    \caption{
        Overview of Tag-Bench.
        Valid scene specifications are selected from a
        schema-derived factor space and realized through
        agentic speech synthesis.
        The resulting synthetic scenes are combined with
        recordings from existing corpora to evaluate
        rich speech annotation under individual and
        composed attribute conditions.
    }
    \label{fig:tag_bench}
\end{figure}

As illustrated in Figure~\ref{fig:tag_bench}, we derive
synthetic scene specifications from selected attributes
of the SURE-Tagger schema.
We first define 11 valid speaker-interaction patterns spanning
speaker count, turn-taking, and overlap.
Combining these patterns with three speaking rates, three SNR
levels, and two settings each for music and reverberation
yields 396 distinct configurations.
We further diversify this controlled set through alternative
dialogue scripts and language variations, including
Chinese--English code-switching, eight additional languages,
and multilingual interactions, to obtain 500 synthetic scenes.
This design supports systematic evaluation of individual
speech attributes and their combinations across diverse conditions.

These scene specifications are then realized as audio through
an agent-orchestrated synthesis pipeline, as illustrated in
Figure~\ref{fig:tag_bench}.
For each scene, the agent coordinates speaker assignments,
turn scheduling, and speech generation, followed by acoustic
rendering under the specified conditions.
The resulting recordings undergo quality checks before
reference annotations are established.

For synthetic speech, reference annotations are derived
from frozen scripts, rendering records, and measurements
of the rendered audio, rather than treating scene
specifications as ground truth.
For existing recordings, we retain source annotations,
supplement missing attributes using Qwen3.5-Omni-Plus~\citep{qwen2026omniplus},
and manually review the annotations.
Attributes that cannot be reliably established remain
\texttt{null}. 

\section{Experiments}
\label{sec:experiments}

\subsection{Rich Speech Annotation Quality}
\label{sec:annotation_quality}

We then evaluate rich speech annotation quality on Tag-Bench.
We compare SURE-Tagger with two audio-understanding
baselines: Qwen3-Omni-Flash~\citep{qwen2025flash}, a general-purpose multimodal
model, and Qwen3-Omni-30B-A3B-Captioner (Qwen-Omni-Captioner)~\citep{qwen2025captioner,xu2025qwen3omni}, a model specialized
for free-form audio descriptions.
Qwen3-Omni-Flash provides observations directly from the
audio, while the Captioner produces a free-form description;
SURE-Tagger instead integrates attribute-specific observations
from specialized speech models.
All three pipelines use the same DeepSeek-V4-Pro model~\citep{deepseek2026v4pro}
to construct the final schema from their respective
observations. The SURE-Tagger observer pipeline contains only \textbf{3.91B}
unique model parameters across its reachable branches,
with 1.95--3.11B involved per recording depending on
the execution route.

We evaluate 19 structured and 10 descriptive axes on both
the existing-corpus and controlled synthetic subsets of
Tag-Bench. For a structured field $f$ in recording $i$,
the score is
\begin{equation}
s_{i,f} =
\begin{cases}
2\operatorname{Sim}_f(\hat v_{i,f},v_{i,f}),
    & \text{valid non-null prediction},\\
1,  & \text{explicit }\texttt{null},\\
0,  & \text{missing, invalid, or failed prediction},
\end{cases}
\label{eq:tagbench_score}
\end{equation}
where $\operatorname{Sim}_f\in[0,1]$ measures field-specific
agreement through exact matching, edit distance, or
numerical error, with class balancing where appropriate.
Descriptive axes use LLM-as-a-judge evaluation~\citep{zheng2023judge}
and are scored by DeepSeek-V4-Pro in $[0,2]$
based on semantic agreement and information coverage
against reference annotations. After normalization to $[0,1]$, we macro-average axis
scores within each reported group (L1, L2, L3, L5, and D).
The overall score equally weights all 29 axes.
For the combined result, axis scores are recomputed
from the pooled existing-corpus and synthetic samples,
rather than averaging the two subset-level scores.
Detailed scoring and aggregation rules are provided in
Appendix~\ref{app:tagbench_scoring}.

\begin{figure*}[t]
    \centering
    \includegraphics[width=\textwidth]{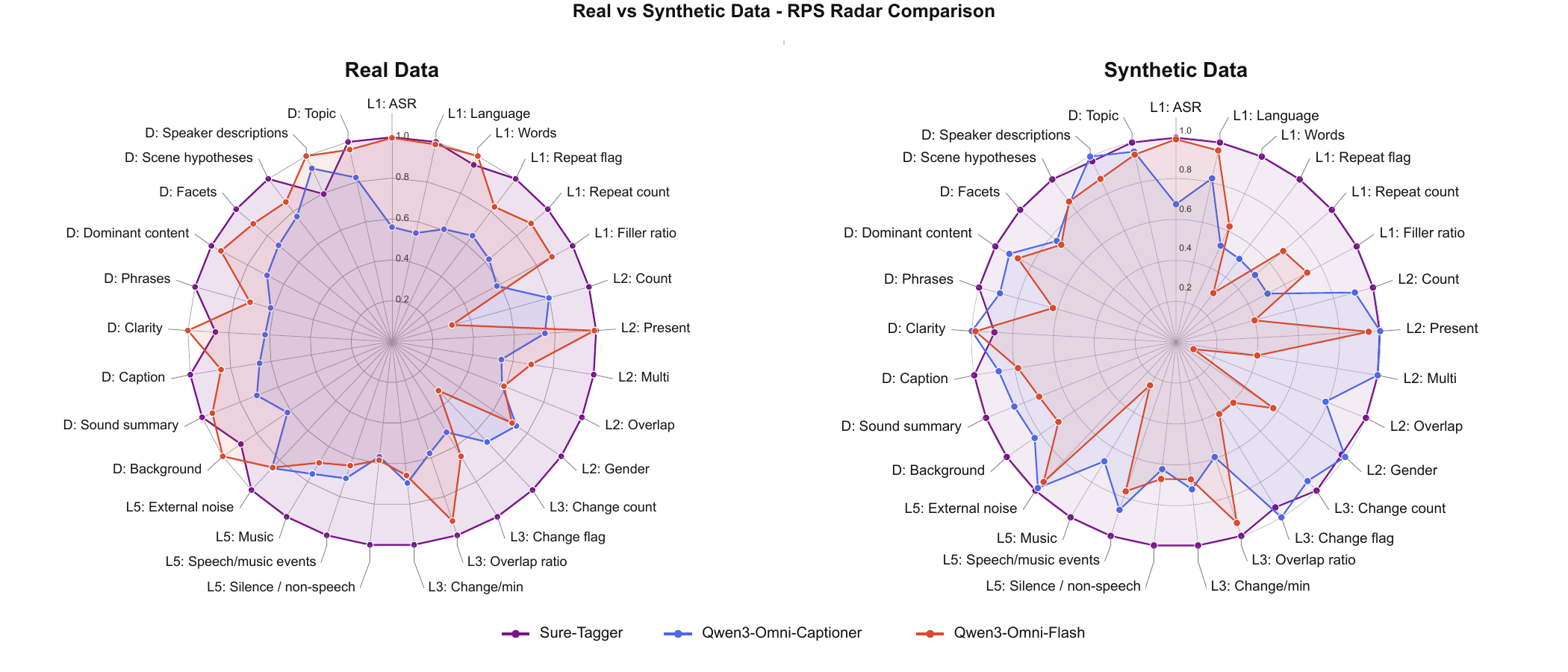}
    \caption{
        Attribute-level annotation performance on Tag-Bench.
        Relative performance scores (RPS) are computed separately
        for existing-corpus and controlled synthetic recordings,
        with the highest-scoring method on each axis normalized to 1.
    }
    \label{fig:tagbench_rps}
\end{figure*}

\begin{table*}[t]
\caption{
Rich speech annotation performance on Tag-Bench.
All methods use the same DeepSeek-V4-Pro language model
for schema construction.
Scores are percentages; higher is better.
Combined scores are recomputed from pooled existing-corpus
and synthetic samples for each axis.
}
\label{tab:tagbench}
\centering
\small
\setlength{\tabcolsep}{4pt}
\renewcommand{\arraystretch}{1.10}

\begin{tabularx}{\textwidth}{
@{}l>{\raggedright\arraybackslash}Xrrrrrrr@{}
}
\toprule
& &
\multicolumn{5}{c}{\textbf{Structured (19)}} &
& \\
\cmidrule(lr){3-7}
\textbf{Subset} &
\textbf{Method} &
\textbf{L1} &
\textbf{L2} &
\textbf{L3} &
\textbf{L5} &
\textbf{Struct.} &
\textbf{D (10)} &
\textbf{Overall (29)} \\
\midrule

Existing
& Qwen3-Omni-Flash
& 79.52 & 55.76 & 53.86 & 55.20
& 62.74 & 62.89 & 62.79 \\

& Qwen3-Omni-Captioner
& 50.29 & 54.00 & 49.74 & 57.64
& 52.70 & 48.44 & 51.23 \\

& \textbf{SURE-Tagger}
& \textbf{84.38} & \textbf{80.62}
& \textbf{81.84} & \textbf{81.07}
& \textbf{82.16} & \textbf{65.09}
& \textbf{76.28} \\

\midrule

Synthetic
& Qwen3-Omni-Flash
& 66.48 & 42.29 & 50.79 & 50.44
& 53.44 & 55.67 & 54.21 \\

& Qwen3-Omni-Captioner
& 55.41 & 75.94 & 66.51 & 63.15
& 64.78 & 61.17 & 63.54 \\

& \textbf{SURE-Tagger}
& \textbf{94.46} & \textbf{79.10}
& \textbf{80.94} & \textbf{82.20}
& \textbf{84.99} & \textbf{67.27}
& \textbf{78.88} \\

\midrule

Combined
& Qwen3-Omni-Flash
& 72.27 & 56.00 & 55.86 & 58.11
& 61.55 & 59.74 & 60.93 \\

& Qwen3-Omni-Captioner
& 52.67 & 65.71 & 59.56 & 66.27
& 60.41 & 53.98 & 58.20 \\

& \textbf{SURE-Tagger}
& \textbf{86.84} & \textbf{81.01}
& \textbf{81.48} & \textbf{82.15}
& \textbf{83.19} & \textbf{66.04}
& \textbf{77.28} \\

\bottomrule
\end{tabularx}

\vspace{2pt}
\begin{minipage}{\textwidth}
\footnotesize
D denotes the descriptive axes.
L4 has no defined scoring axes and is excluded.
The benchmark scores 19 structured and 10 of the
12 descriptive fields.
Combined results are recomputed by pooling samples
within each axis, rather than averaging the Existing
and Synthetic scores.
\end{minipage}
\end{table*}

For attribute-level visualization, we define the relative
performance score (RPS) for method $m$, axis $f$, and
data source $d$ as
\begin{equation}
\operatorname{RPS}_{m,f}^{(d)}
=
\frac{r_{m,f}^{(d)}}
{\max_{m'\in\mathcal M}r_{m',f}^{(d)}},
\label{eq:tagbench_rps}
\end{equation}
where $r_{m,f}^{(d)}$ is the normalized axis score and
$\mathcal M$ is the set of compared methods.
RPS is computed separately for existing-corpus and
synthetic recordings and is used only for the
attribute-level visualization in
Figure~\ref{fig:tagbench_rps}.
Table~\ref{tab:tagbench} reports the absolute scores.
SURE-Tagger achieves 76.28\% on existing-corpus
recordings, compared with 62.79\% for Qwen3-Omni-Flash,
with pronounced gains in speaker, temporal, and
acoustic annotations.
On synthetic recordings, it obtains 78.88\%, compared
with 63.54\% for Qwen3-Omni-Captioner.
Its combined score reaches 77.28\%, versus 60.93\%
and 58.20\% for the two baselines, respectively.

Re-scoring the existing-corpus descriptive axes with
Grok 4.7~\citep{grok2026v47} preserves the method ordering under the
original DeepSeek-V4-Pro judge.
In a separate human study, annotators select
SURE-Tagger in 159 of 200 layer-level judgments
across 20 recordings (Appendix~\ref{app:human_preference})

A language-tag ablation further shows that LLM-based
evidence integration improves structured attribute
prediction beyond schema construction, increasing
the language-axis score from 88.90\% without LLM
integration to 98.62\% with the standard prompt. The Captioner's existing-corpus score near $50\%$ is
dominated by explicit abstention rather than partial
correctness. SURE-Tagger does not rely on this
half-credit floor: after removing explicit
\texttt{null}s, it remains highest overall and on
structured axes (Appendix~\ref{app:null_rates}).


\subsection{Compositional Evaluation of ASR}
\label{sec:compositional_evaluation}

We investigate whether SURE-Tagger enables ASR evaluation
under shared speech conditions across heterogeneous datasets.
We use \citet{openbench} for ASR evaluation and
WER/CER computation.
We sample 100 recordings from each of ten datasets,
yielding 1,000 recordings, and annotate them with SURE-Tagger.
The source corpora are detailed in Appendix~\ref{app:scene_clustering}.
The resulting representations allow us to reorganize
recordings by their speech attributes rather than
their dataset identities.

\begin{figure}[t]
    \centering
    \includegraphics[width=0.8\linewidth]{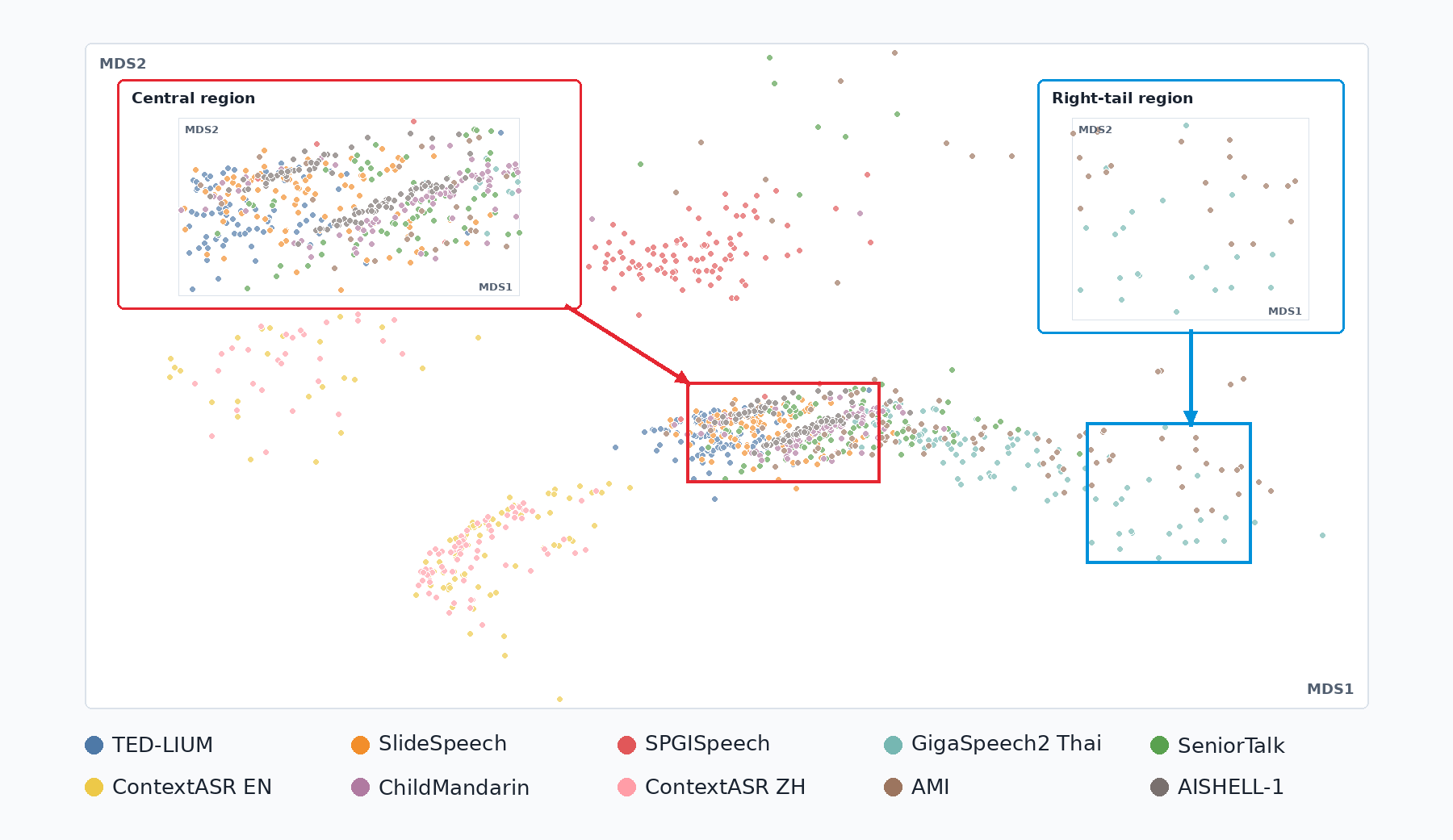}
    \caption{
        Cross-dataset coverage in the schema-derived attribute
        space. Colors indicate source datasets.
        The MDS projection provides a qualitative view of
        within-dataset diversity and shared speech conditions
        across datasets.
    }
    \label{fig:dataset_coverage}
\end{figure}

Figure~\ref{fig:dataset_coverage} visualizes the recordings
in the schema-derived attribute space.
Recordings from the same dataset occupy different regions,
while those from different datasets exhibit overlapping
distributions.
This qualitative view motivates organizing recordings
by speech conditions beyond their dataset identities.
We examine two complementary approaches:
attribute-based clustering to identify recurring conditions,
and explicit conjunctive queries to construct subsets
under user-specified conditions.

To identify recurring speech conditions, we cluster the
recordings using 23 audio-side schema attributes grouped
into four equally weighted categories: content and events,
speaker interaction, acoustic quality, and spatial properties.
The attributes include numerical, Boolean, and set-valued
fields, with missing values retained as unknown.
We define a mixed-type distance as
\begin{equation}
d_{\mathrm{tag}}(x_i,x_j)
=
\frac{1}{|\mathcal A_{ij}|}
\sum_{a\in\mathcal A_{ij}}d_a(x_i,x_j),
\label{eq:tag_distance}
\end{equation}
where $\mathcal A_{ij}$ contains the comparable attribute
groups and $d_a$ accounts for field-specific dissimilarities
and missingness.
We apply k-medoids~\citep{kaufman1990clustering} with attribute-based refinement to obtain
six frozen condition clusters.
Dataset identities, language and transcript features,
and ASR errors are excluded from clustering.
Implementation details are provided in
Appendix~\ref{app:scene_clustering}.

\begin{table}[t]
\caption{
ASR error rates (\%) across six attribute-defined
speech-condition clusters.
Scores are micro-aggregated under the language-specific
WER/CER protocol; lower is better.
}
\label{tab:scene_clusters}
\centering
\small
\setlength{\tabcolsep}{2.5pt}
\renewcommand{\arraystretch}{1.12}

\begin{tabularx}{\linewidth}{
@{}
>{\raggedright\arraybackslash}p{0.21\linewidth}
*{6}{>{\centering\arraybackslash}X}
@{}
}
\toprule
\textbf{ASR model}
& \textbf{C00}
& \textbf{C01}
& \textbf{C02}
& \textbf{C03}
& \textbf{C04}
& \textbf{C05} \\[2pt]

&
\shortstack[c]{Mid-length,\\dry speech}
&
\shortstack[c]{Short,\\low SNR}
&
\shortstack[c]{Short,\\moderate\\reverb}
&
\shortstack[c]{Long, clean\\speech}
&
\shortstack[c]{Very short\\speech-\\sparse}
&
\shortstack[c]{Short, noisy\\with music} \\[3pt]

\midrule

Parakeet-TDT
& 15.47 & 48.06 & 14.84 & 20.18 & 100.00 & 100.00 \\

Qwen3-ASR-1.7B
& 14.70 & 21.53 & 12.59 & 5.12 & 86.36 & 17.26 \\

MOSS-Transcribe
& 12.50 & 36.43 & 12.83 & 15.03 & 85.71 & 91.30 \\

SenseVoiceSmall
& 10.04 & 25.00 & 10.00 & 13.67 & 18.18 & 34.78 \\

Whisper large-v3 turbo
& 17.82 & 51.55 & 18.35 & 17.82 & 104.55 & 130.43 \\

Granite Speech 4.1
& 15.51 & 24.81 & 16.04 & 22.06 & 33.33 & 73.91 \\

FunASR-Flash
& 13.90 & 19.85 & 13.03 & 11.25 & 86.36 & 13.57 \\

\bottomrule
\end{tabularx}

\vspace{2pt}
\begin{minipage}{\linewidth}
\footnotesize
Evaluation coverage varies with model language support.
Cluster sizes and per-model scoring coverage are
provided in Appendix~\ref{app:scene_clustering}.
\end{minipage}
\end{table}

Table~\ref{tab:scene_clusters} reveals distinct error
patterns across the frozen condition clusters.
For example, Qwen3-ASR-1.7B achieves 5.12\% on C03,
characterized by longer, clean speech, compared with
21.53\% on C01, which contains short, low-SNR recordings.
Using the same clustering procedure, caption-derived
attributes yield a weaker association between cluster
membership and recording-level ASR errors
($\eta^2=0.022$) than SURE-Tagger attributes
($\eta^2=0.348$); detailed analyses are provided in
Appendix~\ref{app:scene_clustering}.
These condition-level error profiles provide a basis
for the targeted data selection investigated in
Section~\ref{sec:targeted_selection}.

\begin{figure}[t]
    \centering
    \includegraphics[width=\linewidth]{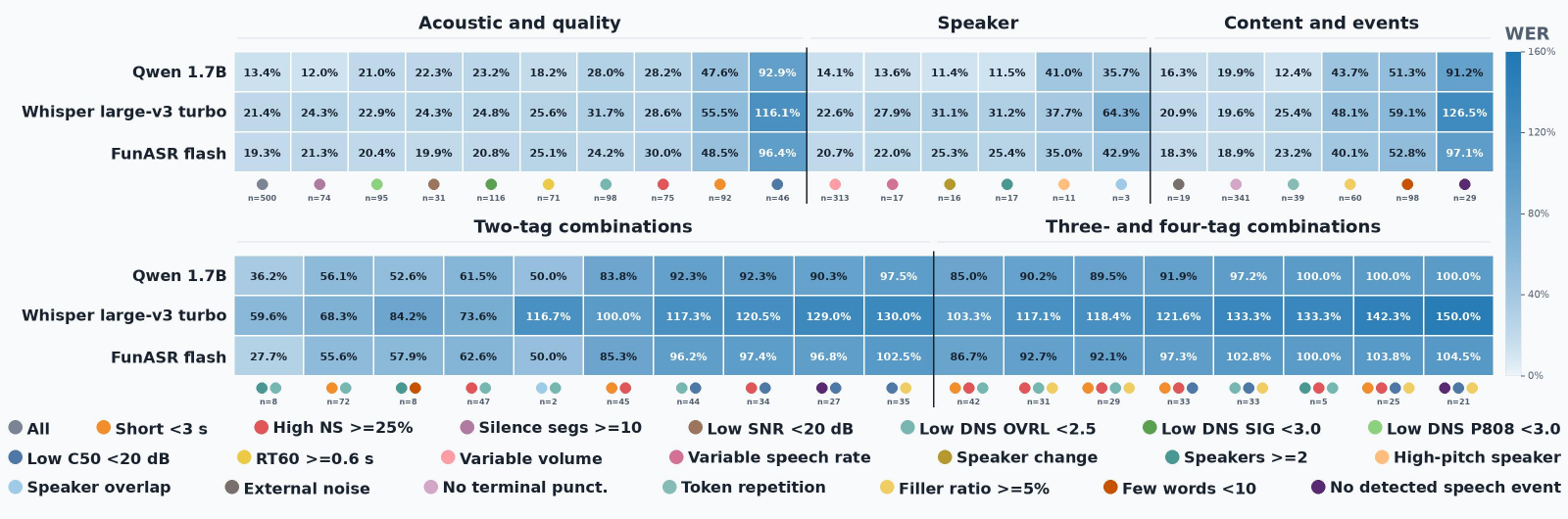}
    \caption{
        Condition-wise ASR performance under individual
        and composed speech conditions.
        WER is computed over reference units within each
        subset using a common 500-recording cohort.
        Sample counts are shown below the condition labels.
        Results for small subsets are exploratory.
    }
    \label{fig:condition_wer}
\end{figure}

Beyond clustering, explicit schema values support
evaluation subsets defined by conjunctions of
speech-attribute predicates.
Following Section~\ref{sec:sure_tagger}, we construct
cross-dataset subsets under individual and composed
conditions and compute corpus-level WER within each subset.
Figure~\ref{fig:condition_wer} compares three ASR models
on a common subset of 500 English recordings, ensuring
that their results are computed over the same samples.

Composing attributes yields more specific evaluation
subsets and exposes differences that may be obscured
by individual-condition results.
For instance, Qwen3-ASR-1.7B obtains 47.6\% WER on
short recordings, compared with 83.8\% on the subset
that additionally has a high non-speech proportion.
Similar patterns are observed for the other two models.
These comparisons describe performance on nested,
partly overlapping subsets and do not isolate the
causal effect of any single attribute.
Together, attribute-defined clusters provide reusable
condition-level error profiles, while conjunctive queries
enable targeted evaluation under specified combinations
of speech attributes.

\subsection{Targeted Data Selection}
\label{sec:targeted_selection}

\begin{table}[t]
\caption{
Targeted data selection and fine-tuning results under
an approximately one-hour audio budget.
(a) Recognition error after fine-tuning, reported as
mean $\pm$ standard deviation over three sampling seeds.
(b) Shared budgeted selection procedure.
Relative change ($\Delta$) is computed against the
unadapted model.
WER and relative change ($\Delta$) are reported in percentages.
}
\label{tab:targeted_selection}
\centering
\small

\begin{minipage}[t]{0.45\linewidth}
\vspace{0pt}
\textbf{(a) Fine-tuning results}
\par\vspace{0pt}

\setlength{\tabcolsep}{2pt}
\renewcommand{\arraystretch}{1.00}

\begin{tabularx}{\linewidth}{
    @{}
    >{\raggedright\arraybackslash}X
    >{\centering\arraybackslash}p{0.40\linewidth}
    >{\raggedleft\arraybackslash}p{0.16\linewidth}
    @{}
}
\toprule
\textbf{Selection}
& \shortstack[c]{\textbf{Error (\%)}\\\textbf{mean $\pm$ SD}}
& \textbf{$\Delta$} \\
\midrule

Original
& 5.50 & -- \\

Random
& $6.17\pm0.34$ & $+12.3$ \\

\addlinespace[3pt]
\midrule

Pool / SURE
& $\mathbf{4.71}\pm\mathbf{0.08}$
& $\mathbf{-14.3}$ \\

Pool / Captioner
& $6.17\pm1.02$ & $+12.4$ \\

\addlinespace[3pt]
\midrule

Eval. / SURE
& $4.96\pm0.27$ & $-9.7$ \\

Eval. / Captioner
& $6.83\pm0.61$ & $+24.4$ \\

\bottomrule
\end{tabularx}
\end{minipage}%
\hfill
\vrule width 0.4pt
\hfill
\begin{minipage}[t]{0.52\linewidth}
\vspace{0pt}
\textbf{(b) Budgeted selection}
\par\vspace{0pt}

\setlength{\tabcolsep}{1.5pt}
\renewcommand{\arraystretch}{1.00}

\begin{tabularx}{\linewidth}{
    @{}r@{\hspace{4pt}}
    >{\raggedright\arraybackslash}X@{}
}
\toprule

\multicolumn{2}{@{}l}{
    \textbf{Input:}
    $\mathcal P$, $R$,
    $s\in\{\mathrm{Eval},\mathrm{Pool}\}$, $B$.
} \\

\midrule

1 &
\textbf{if} $s=\mathrm{Eval}$:
reuse evaluation clusters and errors \\
&
\quad $(M,r)\gets
(M_{\mathcal E}^{R},r_{\mathcal E}^{R})$ \\[2pt]

2 &
\textbf{else (Pool):}
cluster and score the candidate pool \\
&
\quad $(M,r)\gets
\operatorname{ClusterScore}(\mathcal P,R)$ \\

\midrule

3 &
\textbf{for} $x_i\in\mathcal P$: \\

4 &
\quad
$g_i\gets\arg\min_g d(R(x_i),m_g)$ \\

5 &
\quad
$w_i\gets(r_{g_i}+10^{-3})^2$ \\[2pt]

6 &
$\mathcal S\gets
\operatorname{WRS}_{B}(\mathcal P,\{w_i\})$ \\

\midrule

\multicolumn{2}{@{}l}{
    \textbf{Output:}
    selected subset $\mathcal S$.
} \\

\bottomrule
\end{tabularx}
\end{minipage}

\par\vspace{0pt}

\begin{minipage}{\linewidth}
\footnotesize
$\mathcal P$ is the candidate pool;
$\mathcal E$ is the evaluation cohort; and $R$
denotes SURE-Tagger or caption-derived attributes.
$M=\{m_g\}$ and $r=\{r_g\}$ are cluster medoids
and their ASR error rates.
$\operatorname{ClusterScore}$ clusters the candidate
pool and estimates its cluster errors using the
original ASR model.
The distance $d$ follows
Section~\ref{sec:compositional_evaluation};
$\operatorname{WRS}_{B}$ samples without replacement
under $B=3600$ s.
All models are evaluated on the same 1,000 recordings.
Overall error pools English WER and Chinese/Thai CER
by reference units.
\end{minipage}

\end{table}

We examine whether condition-level ASR errors can
guide fine-tuning data selection.
From the same ten sources, we construct a pool of
10,000 candidate recordings, disjoint from the
1,000-recording evaluation cohort.
We select approximately one hour of audio to
fine-tune Qwen3-ASR-1.7B~\citep{shi2026qwen3asr}.

Both strategies assign candidates to speech-condition
clusters and sample them according to cluster-level,
rather than individual, ASR errors.
In \emph{evaluation-cluster-guided selection} (Eval.),
candidates inherit error rates from the frozen
evaluation clusters of
Section~\ref{sec:compositional_evaluation}.
In \emph{candidate-pool-guided selection} (Pool),
we instead cluster the candidate pool and estimate
cluster errors using the original ASR model.
Table~\ref{tab:targeted_selection}(b) summarizes
the shared selection procedure.

For each strategy, we compare SURE-Tagger with
caption-derived attributes, reconstructing the
corresponding clusters while keeping the selection
rule and fine-tuning settings unchanged.
We also include random selection and the unadapted
model as baselines.
Selection and fine-tuning are repeated with three
sampling seeds.
All models are evaluated on the same cohort;
Eval. uses this cohort's errors for selection,
whereas Pool does not.

Table~\ref{tab:targeted_selection}(a) shows that
Pool-guided selection with SURE-Tagger reduces
overall error from 5.50\% to
$(4.71\pm0.08)\%$, a 14.3\% relative reduction.
Under the same strategy, caption-derived attributes
yield $(6.17\pm1.02)\%$, compared with
$(6.17\pm0.34)\%$ for random selection.
Eval.-guided selection yields $(4.96\pm0.27)\%$
with SURE-Tagger and $(6.83\pm0.61)\%$
with caption-derived attributes.
Both SURE-Tagger variants improve upon the
unadapted model in all three runs, supporting
the utility of its representations for
condition-guided selection in this setting.
We additionally adapt Whisper-small~\citep{radford2023whisper} under the same
one-hour budget; results are reported in
Appendix~\ref{app:whisper_selection}.

\section{Conclusion}

We presented SURE-Tagger, a framework for constructing
semi-structured speech representations with explicit
attributes, uncertainty states, and supplementary
open-ended descriptions.
Experiments on Tag-Bench demonstrate improved annotation
quality over audio-understanding and captioning baselines.
Beyond annotation, these representations support
cross-dataset compositional evaluation and
condition-guided data selection.
Under a one-hour fine-tuning budget, candidate-pool-guided
selection using SURE-Tagger reduces the overall ASR
error rate of Qwen3-ASR-1.7B from 5.50\% to
$(4.71\pm0.08)\%$ across three sampling seeds. Future work will examine broader generalization across
models, languages, and recording conditions, improve
uncertainty handling, and reduce the computational cost
of candidate-pool data selection.

\newpage
\subsection*{AI use statement}

In this work, generative AI tools were used as components of the proposed
speech annotation and evaluation pipeline. Specifically, large language models
were used to integrate observations from specialized speech models into the
structured speech representation, to assist with the construction and
evaluation of synthetic speech scenes, and to score open-ended descriptive
fields. Generative speech tools were also used to create the controlled
synthetic speech subset described in Section \ref{sec:tag_bench} and Appendix \ref{app:tag_bench_details}.
We did not use generative AI tools to replace the authors' responsibility for
the research questions, experimental design, analysis, or scientific claims.
All generated data, model outputs, annotations, and evaluation results were
checked using the validation, post-render verification, and manual review
procedures described in the paper and supplementary material. We reviewed all
AI-assisted content and take full responsibility for the final manuscript,
methods, results, and conclusions.

\subsection*{Ethics statement}

This work uses speech recordings from publicly available datasets and
controlled synthetic speech scenes. We follow the licenses and usage
conditions of the source datasets and do not release restricted source audio
or other materials beyond their permitted scope. Synthetic scenes are
generated for controlled evaluation and are not intended to imitate,
identify, or impersonate specific individuals.
The human preference study described in Appendix~B uses speech recordings for
comparative evaluation of system outputs. Annotators were instructed to
evaluate only the specified attribute layer, and system outputs were presented
in a randomized order. The study does not collect new personal information or
infer sensitive personal attributes from participants. We report only
aggregate judgments and do not disclose annotator identities.
Some source datasets contain demographic or sensitive speech-related
attributes. These attributes are treated as dataset-provided annotations or
evaluation labels rather than reliable properties inferred from speech. We
retain explicit uncertainty when the available evidence is insufficient and
do not treat missing information as confirmed absence. The authors take
responsibility for complying with applicable data licenses, privacy
requirements, and research ethics policies.

\subsection*{Reproducibility statement}

The paper and supplementary material provide the definitions of the
semi-structured speech representation, the Tag-Bench construction procedure,
the annotation scoring rules, the clustering and data-selection algorithms,
the synthetic speech rendering settings, and the additional robustness
experiments. The complete implementation, configuration files, prompts,
evaluation scripts, dataset manifests, and derived annotations will be
released as open-source materials upon publication, subject to the licenses
and redistribution restrictions of the source datasets. Restricted audio
resources will be referenced through their original dataset sources, while
the released package will include reproducible download, preprocessing, and
evaluation scripts whenever redistribution of the audio is not permitted.
The released code and data package will also include the random seeds, model
and checkpoint specifications, training and decoding configurations, sample
splits, and scripts for reproducing the main tables and figures. We will
provide an anonymized version of the reproducibility materials during review
when required by the double-blind submission policy.
\bibliographystyle{iclr2027_conference}
\bibliography{reference}
\appendix
\section{Tag-Bench Scoring Details}
\label{app:tagbench_scoring}

This appendix specifies the 29 evaluation axes used in
Section~\ref{sec:experiments} and the field-specific metrics
that instantiate $\operatorname{Sim}_f$ in
Equation~\ref{eq:tagbench_score}.
The axes follow the benchmark-v1 layering:
L1 lexical content (6),
L2 speaker (5),
L3 temporal structure (4),
L5 acoustic scenes and events (4),
and an additional descriptive layer D (10).
Layer~4 (paralinguistic and non-verbal information) has no
frozen reference and is not scored.
An axis is scored on a recording only when a valid reference
is available; undefined scores are excluded rather than
recorded as zero.
The same operators are applied to existing-corpus and
synthetic recordings.

\subsection{Prediction States and Valid Samples}
For recording $i$ and field $f$, a prediction is first
normalized into one of five states:
\texttt{known} (present and schema-valid),
\texttt{null} (explicit abstention),
\texttt{missing},
\texttt{schema-invalid}, or
\texttt{runtime-failed}.
The per-sample score $s_{i,f}\in[0,2]$ follows
Equation~\ref{eq:tagbench_score}:
a valid non-null prediction receives
$2\operatorname{Sim}_f(\hat v_{i,f},v_{i,f})$,
an explicit \texttt{null} receives $1$,
and all other states receive $0$.
Thus a score near $50\%$ can arise from systematic
abstention and does not imply $50\%$ identification
accuracy.

A sample contributes to axis $f$ only if the reference is
available, schema-legal, and free of input leakage.
Let $\mathcal{S}_{f,g}$ denote the valid samples for field
$f$ in evaluation group $g$, and
$N_{f,g}=\lvert\mathcal{S}_{f,g}\rvert$.
If $N_{f,g}=0$, group $g$ does not contribute to axis $f$.

Unless otherwise specified, the group score is the micro-average
\begin{equation}
r_{f,g}
=
\frac{\sum_{i\in\mathcal{S}_{f,g}} s_{i,f}}{2N_{f,g}}
\in[0,1].
\end{equation}
Boolean, multi-label, language, ASR, and lexical axes replace
this micro-average with the class- or language-balanced rules
below.
All reported percentages are $100r_f$.

\subsection{Scoring Operators}
Write $p$ for a known prediction and $g$ for the reference.
The structured operators are:
\begin{align}
s^{\mathrm{eq}}(p,g)
&=
\begin{cases}
2, & p=g,\\
0, & p\neq g,
\end{cases}
\label{eq:app_eq}\\
s^{\mathrm{abs}}(p,g)
&=
2\bigl(1-\lvert p-g\rvert\bigr),
\label{eq:app_abs}\\
s^{\mathrm{cont}}(p,g;T)
&=
2\max\bigl(0,\,1-\lvert p-g\rvert/T\bigr),
\label{eq:app_cont}\\
s^{\mathrm{nae}}(p,g)
&=
2\left(1-\frac{\lvert p-g\rvert}{\max(p,g,1)}\right).
\label{eq:app_nae}
\end{align}
Equation~\ref{eq:app_abs} is used for ratios in $[0,1]$;
Equation~\ref{eq:app_cont} uses tolerance $T>0$;
Equation~\ref{eq:app_nae} is a symmetric normalized absolute
error for counts and rates.
In all cases $\operatorname{Sim}_f=s/2$.
Explicit \texttt{null} still scores $1$ and is not passed
through these operators.

\paragraph{Class-balanced Boolean aggregation.}
Let $\mathcal{S}^+$ and $\mathcal{S}^-$ be the positive and
negative reference sets, with $N^\pm=\lvert\mathcal{S}^\pm\rvert$,
and let $r^\pm$ be the corresponding micro-averages of $s_i/2$.
The group score is
$(r^++r^-)/2$ when both classes are present,
$r^+$ when only positives are present,
and \emph{undefined} when $N^+=0$.
A class with no positive examples is used only to audit false
positives and does not enter the 29-axis denominator.

\paragraph{Multi-label macro-average.}
For a fixed label set $\mathcal{C}$, each class $c$ is scored
as a Boolean axis.
The group score averages $r_c$ over classes with at least one
positive example.

\paragraph{Language-grouped macro-average.}
Language-dependent axes first average within each reference
language, then average languages with equal weight.
ASR further groups by language and by WER versus CER.

\subsection{Layered Axis Inventory}
Table~\ref{tab:app_layers} summarizes the scored layers.
Tables~\ref{tab:app_struct_axes} and~\ref{tab:app_desc_axes}
list the 19 structured axes (L1, L2, L3, L5) and the 10
descriptive axes (D).
The overall score is an equal-weight average of all 29 axes.

\begin{table}[t]
\caption{Tag-Bench scoring layers, aligned with the
benchmark-v1 report.
Layer~4 is recorded as n/a and is not among the 29 axes.}
\label{tab:app_layers}
\centering
\footnotesize
\setlength{\tabcolsep}{4pt}
\renewcommand{\arraystretch}{1.08}
\begin{tabularx}{\linewidth}{@{}llc >{\raggedright\arraybackslash}X@{}}
\toprule
Layer & Name & \# & Scored axes \\
\midrule
L1 & Lexical content & 6
  & ASR transcript, language, word count, repetition presence,
    repetition count, filler ratio \\
L2 & Speaker & 5
  & Speaker count, speaker present, multi-speaker,
    speaker overlap, gender sequence \\
L3 & Temporal structure & 4
  & Speaker-change count, speaker change, overlap ratio,
    speaker-change rate \\
L4 & Paralinguistic / non-verbal & 0
  & No frozen reference; excluded \\
L5 & Acoustic scenes and events & 4
  & Non-speech ratio, speech/music events, music present,
    external noise type \\
D & Descriptive labels & 10
  & Caption, topic, background condition, content impression,
    clarity/intelligibility, background sound summary,
    descriptive phrases, facets, speaker descriptions,
    scene hypotheses \\
\midrule
T & All labels & 29
  & 19 structured + 10 descriptive, equal-axis average \\
\bottomrule
\end{tabularx}
\end{table}

\begin{table}[t]
\caption{Nineteen structured axes on Tag-Bench (L1, L2, L3, L5).
\texttt{T1}: AMI meeting speech~\citep{carletta2006ami};
\texttt{T2}: FLEURS~\citep{conneau2022fleurs} and LibriSpeech~\citep{panayotov2015librispeech};
\texttt{T3}: DESED~\citep{turpault2019desed} and FSD50K~\citep{fonseca2020fsd50k}.
}
\label{tab:app_struct_axes}
\centering
\footnotesize
\setlength{\tabcolsep}{3pt}
\renewcommand{\arraystretch}{1.08}
\begin{tabularx}{\linewidth}{@{}rlll >{\raggedright\arraybackslash}X@{}}
\toprule
\# & Layer & Axis & Scope & Known-case metric \\
\midrule
1 & L1 & ASR transcript
  & T1+T2
  & $1-\min(d/\max(\lvert r\rvert,1),1)$ \\
2 & L1 & Language
  & T1+T2
  & Exact match on 10-language set \\
3 & L1 & Word count
  & T1+T2
  & Symmetric NAE \\
4 & L1 & Has repetition
  & T1+T2
  & Boolean match, class-balanced \\
5 & L1 & Repetition count
  & T1+T2
  & Exact integer match \\
6 & L1 & Filler ratio
  & English, word count${}>0$
  & Continuous score, $T=0.05$ \\
7 & L2 & Speaker count
  & T1
  & Exact integer match \\
8 & L2 & Speaker present
  & T1
  & Boolean match, class-balanced \\
9 & L2 & Multi-speaker
  & T1
  & Boolean match, class-balanced \\
10 & L2 & Speaker overlap
  & T1
  & Boolean match, class-balanced \\
11 & L2 & Gender sequence
  & T1
  & $1-d/K$ on first-speak order \\
12 & L3 & Speaker-change count
  & T1
  & Symmetric NAE \\
13 & L3 & Speaker change
  & T1
  & Boolean match, class-balanced \\
14 & L3 & Overlap ratio
  & T1
  & Absolute ratio error \\
15 & L3 & Speaker-change rate
  & T1, $D>0$
  & Symmetric NAE \\
16 & L5 & Non-speech ratio
  & T1
  & Absolute ratio error \\
17 & L5 & Speech/music events
  & T3
  & Per-class Boolean macro-average \\
18 & L5 & Music present
  & T3 FSD50K
  & Boolean match, class-balanced \\
19 & L5 & External noise type
  & T3
  & Per-class Boolean macro-average \\
\bottomrule
\end{tabularx}
\end{table}

\begin{table}[t]
\caption{Ten descriptive axes in layer D, scored by
DeepSeek-V4-Pro as continuous values $u\in[0,2]$ and
reported as $r=\bar u/2$.}
\label{tab:app_desc_axes}
\centering
\footnotesize
\setlength{\tabcolsep}{3pt}
\renewcommand{\arraystretch}{1.08}
\begin{tabularx}{\linewidth}{@{}rll >{\raggedright\arraybackslash}X@{}}
\toprule
\# & Layer & Axis & Comparison target \\
\midrule
20 & D & Caption
  & Holistic caption semantics and coverage \\
21 & D & Topic
  & Main topic / activity \\
22 & D & Background condition
  & Acoustic background and scene condition \\
23 & D & Dominant content impression
  & Salient content percept \\
24 & D & Clarity / intelligibility
  & Perceived clarity and intelligibility \\
25 & D & Background sound summary
  & Summary of non-speech sounds \\
26 & D & Descriptive phrases
  & Unordered set of short phrases \\
27 & D & Facets
  & Extensible named descriptive facets \\
28 & D & Speaker descriptions
  & Per-speaker natural-language descriptions \\
29 & D & Scene hypotheses
  & Cautious scene / interaction hypotheses \\
\bottomrule
\end{tabularx}
\end{table}

\subsection{L1 Lexical Content}
Transcripts are normalized by Unicode NFKC, lowercasing,
punctuation stripping, and whitespace folding.
Tokenization is character-level for Chinese, Japanese, and
Korean, and whitespace-delimited otherwise.
Adjacent-repetition counts greedily scan token $n$-grams for
$n\in\{3,2,1\}$.
English filler words are
\{\texttt{ah}, \texttt{er}, \texttt{erm}, \texttt{hmm},
\texttt{mm}, \texttt{oh}, \texttt{uh}, \texttt{um},
\texttt{yeah}\}; other languages are not scored for filler
ratio.

\paragraph{ASR transcript.}
Let $\mathbf{r}$ and $\mathbf{h}$ be the tokenized reference
and hypothesis, and let $d$ be their Levenshtein distance
(unit insert/delete/substitute cost).
Then
\begin{equation}
\operatorname{Sim}
=
1-\min\bigl(d/\max(\lvert\mathbf{r}\rvert,1),\,1\bigr).
\end{equation}
Chinese, Japanese, and Korean use CER; all other languages
use WER.
Group scores first average within
$(\mathrm{language},\mathrm{WER/CER})$ bins, then average
bins and evaluation groups with equal weight.

\paragraph{Language.}
The closed set is
$\{\mathrm{ar},\mathrm{zh},\mathrm{en},\mathrm{es},\mathrm{fr},\mathrm{hi},\mathrm{ja},\mathrm{ko},\mathrm{ru},\mathrm{sw}\}$.
Known predictions score $2$ iff the normalized language code
matches.
Scores are language-macro-averaged within each group, then
averaged across T1, FLEURS, and LibriSpeech.

\paragraph{Word count, repetition, and filler ratio.}
Word count uses $s^{\mathrm{nae}}$.
\texttt{has\_repetition} is a class-balanced Boolean of
whether the adjacent-repetition count is positive;
\texttt{repetition\_count} is an exact integer match.
Filler ratio uses $s^{\mathrm{cont}}(\cdot,\cdot;0.05)$ on
English clips with positive word count.
These lexical axes are language-grouped and then averaged
across T1/T2 groups.

\subsection{L2 Speaker}
Speaker-timeline references on AMI are obtained by merging
per-speaker utterance turns, taking the speech union $U$,
the overlap duration $O$ of simultaneously active speakers
with a $0.1\,\mathrm{s}$ collar, and the speaker-change count
$C$ (a change is recorded when adjacent turns have different
speakers and a gap of at most $1.0\,\mathrm{s}$).
Background voices such as television or radio are not counted
as conversational speakers.
Let $K$ be the number of distinct speakers and $D$ the clip
duration.
Then non-speech ratio is $\mathrm{clip}_{[0,1]}(1-U/D)$,
overlap ratio is $O/U$ when $U>0$, and speaker-change rate
is $60C/D$ when $D>0$.

\paragraph{Speaker inventory.}
\texttt{speaker\_count} is an exact match to $K$.
\texttt{speaker\_present}, \texttt{multi\_speaker}, and
\texttt{speaker\_overlap} are class-balanced Booleans for
$K>0$, $K>1$, and $O\ge 0.1\,\mathrm{s}$, respectively.
These four axes are scored independently; cross-field
consistency is audited but does not override the axis scores.

\paragraph{Gender sequence.}
On AMI, gender is a sequence
$\mathbf{g}=(g_1,\ldots,g_K)$ in first-speak order, with
$g_j\in\{\mathrm{male},\mathrm{female}\}$.
For a known prediction $\mathbf{p}$ of length $M$,
$d=\operatorname{Lev}(\mathbf{g},\mathbf{p})$ and
\begin{equation}
s
=
\begin{cases}
2\max(0,1-d/K), & K>0,\\
2\cdot\mathbf{1}[M=0], & K=0.
\end{cases}
\end{equation}
Gender must not be inferred from pitch, timbre, or speaking
rate in the scoring protocol; the axis uses the frozen AMI
speaker-gender reference.

\subsection{L3 Temporal Structure}
\texttt{speaker\_change\_count} and
\texttt{speaker\_change\_rate\_per\_min} use $s^{\mathrm{nae}}$.
\texttt{speaker\_change} is the class-balanced Boolean $C>0$.
\texttt{overlap\_ratio} uses $s^{\mathrm{abs}}$.
All four L3 axes are scored on T1.

\subsection{L5 Acoustic Scenes and Events}
\paragraph{Non-speech ratio.}
This axis measures the fraction of time without detected
human speech, including music and environmental sound, and
uses $s^{\mathrm{abs}}$ on T1.

\paragraph{Events and noise.}
On T3, \texttt{speech\_music\_events} is a multi-label set
over $\{\mathrm{speech},\mathrm{music},\mathrm{singing}\}$,
and \texttt{external\_noise\_type} is a multi-label set over
$\{\mathrm{music},\mathrm{animal},\mathrm{mechanical},\mathrm{nature},\mathrm{formless},\mathrm{channel\_environment}\}$.
Native source labels are mapped into this closed world;
unmapped target classes are treated as negatives for
evaluation.
Classes without positive examples, including
\texttt{singing} and \texttt{channel\_environment} in the
current subset, do not enter the macro-average.
\texttt{music\_present} is a class-balanced Boolean and is
scored only on FSD50K, because the DESED split contains no
positive music examples.
An empty array \texttt{[]} asserts that no listed class is
present and is not equivalent to \texttt{null}.

\subsection{D Descriptive Labels}
Descriptive fields are open natural language and are not
scored with the structured $2/1/0$ operators.
DeepSeek-V4-Pro~\citep{deepseek2026v4pro} assigns a continuous score
$u_{i,f}\in[0,2]$ to each field from the pair of reference
and predicted values.
An explicit predicted \texttt{null} is fixed at $u=1.00$.
The remaining anchors are: $2.00$ for equivalent central
proposition and complete coverage;
$1.70$--$1.99$ for near-equivalent predictions with only
minor omissions;
$1.25$--$1.69$ for a shared core with important missing,
over-broad, or weakly supported details;
$1.01$--$1.24$ for weak relatedness;
$0.01$--$0.99$ for a largely incorrect core with residual
valid information;
and $0.00$ for unrelated or contradictory content.
Paraphrase, translation, and harmless wording differences
are not penalized.

The judge compares semantic agreement and information
coverage, weighting the main object over secondary
attributes.
Array-valued fields are compared as unordered semantic sets:
\texttt{descriptive\_phrases} by concept recall/precision,
\texttt{facets} by the joint role of name and description
rather than literal facet names, and
\texttt{scene\_hypotheses} by compatible underlying
possibilities rather than confidence metadata.
\texttt{speaker\_descriptions} are aligned by speaker
identity when IDs are present; omitting a speaker or
attaching a material attribute to the wrong speaker is
penalized.
Speaker gender in these descriptions is ignored here,
because it is scored as structured axis~11 in L2.
Auxiliary fields such as \texttt{uncertainties} and
\texttt{usability\_notes} are not scored.

Let $\mathcal{S}_f^{\mathrm{ok}}$ be the samples with a
successful judge output.
The axis score is
\begin{equation}
r_f
=
\frac{1}{2\lvert\mathcal{S}_f^{\mathrm{ok}}\rvert}
\sum_{i\in\mathcal{S}_f^{\mathrm{ok}}} u_{i,f}.
\end{equation}
Failed or missing descriptive judgments are excluded rather
than scored as zero.

\subsection{Aggregation}
Structured axis scores use label-first macro-averaging:
a field is first averaged with equal weight over the
evaluation groups on which it is defined, then layer scores
average fields with equal weight.
Descriptive axes enter as $r_f=\bar u_f/2$.
Let $\mathcal{F}_{\mathrm{L1}}$,
$\mathcal{F}_{\mathrm{L2}}$,
$\mathcal{F}_{\mathrm{L3}}$,
$\mathcal{F}_{\mathrm{L5}}$, and
$\mathcal{F}_{\mathrm{D}}$ denote the scored axes in each
layer.
Then
\begin{equation}
\begin{aligned}
R_{\mathrm{L1}}
&=
\frac{1}{6}\sum_{f\in\mathcal{F}_{\mathrm{L1}}} r_f,
\quad
R_{\mathrm{L2}}
=
\frac{1}{5}\sum_{f\in\mathcal{F}_{\mathrm{L2}}} r_f,\\
R_{\mathrm{L3}}
&=
\frac{1}{4}\sum_{f\in\mathcal{F}_{\mathrm{L3}}} r_f,
\quad
R_{\mathrm{L5}}
=
\frac{1}{4}\sum_{f\in\mathcal{F}_{\mathrm{L5}}} r_f,\\
R_{\mathrm{D}}
&=
\frac{1}{10}\sum_{f\in\mathcal{F}_{\mathrm{D}}} r_f,
\quad
R_{\mathrm{T}}
=
\frac{1}{29}\sum_{f\in\mathcal{F}_{29}} r_f.
\end{aligned}
\end{equation}
These aggregation rules produce the layer-level and
overall scores reported in Table~\ref{tab:tagbench},
with D included as an additional scored layer.
We do not average over the Cartesian product of groups and
fields in one step, so an axis defined on a single group
(e.g., gender on T1, music presence on FSD50K) is not
diluted by undefined cells.

Relative performance scores in
Figure~\ref{fig:tagbench_rps} are computed separately for
existing-corpus and synthetic recordings by dividing each
axis score by the best method on that axis and source.
RPS is used only for visualization; Table~\ref{tab:tagbench}
reports absolute equal-axis percentages.

\section{Judge Robustness and Human Preference}
\label{app:second_judge}

This appendix reports two checks of the Tag-Bench protocol.
The first re-scores the descriptive axes with a second LLM
judge.
The second asks human listeners to pick the best system per
layer under the same present-and-correct $>$ absent $>$
incorrect ordering used in Equation~\ref{eq:tagbench_score}.

\subsection{Second-Judge Robustness of Descriptive Scoring}

Table~\ref{tab:tagbench} reports descriptive scores under
DeepSeek-V4-Pro~\citep{deepseek2026v4pro}.
Because these ten axes are open natural language, the
absolute percentages can depend on the judge.
We re-score the \emph{same} existing-corpus
predictions with Grok~4.7~\citep{grok2026v47}, holding the prompt, references,
field definitions, and aggregation rule fixed
(Section~\ref{app:tagbench_scoring}).
Schema generation is not rerun: only the descriptive judge
changes.
Official Tag-Bench numbers remain those of DeepSeek-V4-Pro;
Grok~4.7 is used only as a second judge.

The comparison uses the 650 existing-corpus recordings.
Two clips lack descriptive references and are excluded for
both judges, leaving 648 scored samples.
Failed or missing judgments are dropped rather than scored
as zero, as in the main protocol.
The synthetic subset is not re-scored here.

\begin{table}[t]
\caption{Existing-corpus descriptive macro-average
($R_{\mathrm{D}}$, 10 axes) under two judges.
$\Delta$ is Grok~4.7 minus DeepSeek-V4-Pro, in percentage
points.
The method ranking is unchanged.}
\label{tab:app_second_judge_macro}
\centering
\footnotesize
\setlength{\tabcolsep}{4pt}
\renewcommand{\arraystretch}{1.12}
\begin{tabular*}{\linewidth}{@{\extracolsep{\fill}}lrrr}
\toprule
Method
  & DeepSeek-V4-Pro
  & Grok~4.7
  & $\Delta$ \\
\midrule
SURE-Tagger
  & 65.09 & \textbf{66.45} & $+1.35$ \\
Qwen3-Omni-Flash
  & 62.89 & 64.90 & $+2.01$ \\
Qwen3-Omni-Captioner
  & 48.44 & 50.72 & $+2.29$ \\
\bottomrule
\end{tabular*}
\end{table}

Table~\ref{tab:app_second_judge_macro} shows that Grok~4.7
is slightly more lenient on all three methods
($+1.35$ to $+2.29$ percentage points), but the ranking
is identical to Table~\ref{tab:tagbench}:
SURE-Tagger $>$ Qwen3-Omni-Flash $>$ Qwen3-Omni-Captioner.
The DeepSeek--Grok gap is smaller for SURE-Tagger than for
the two caption-based systems.

\begin{table}[t]
\caption{Per-axis existing-corpus scores under Grok~4.7,
with the change relative to DeepSeek-V4-Pro (percentage
points).
Axis names follow Table~\ref{tab:app_desc_axes}.}
\label{tab:app_second_judge_axes}
\centering
\footnotesize
\setlength{\tabcolsep}{2.5pt}
\renewcommand{\arraystretch}{1.08}
\begin{tabular*}{\linewidth}{@{\extracolsep{\fill}}l rrr rrr}
\toprule
 & \multicolumn{3}{c}{Grok~4.7}
 & \multicolumn{3}{c}{$\Delta$} \\
\cmidrule(lr){2-4}\cmidrule(lr){5-7}
Axis
  & SURE
  & Omni
  & Cap.
  & SURE
  & Omni
  & Cap. \\
\midrule
Caption
  & 68.49 & 63.50 & 50.41
  & $-0.78$ & $+4.81$ & $+4.94$ \\
Topic
  & 79.36 & 75.99 & 67.29
  & $-0.22$ & $-0.53$ & $+1.90$ \\
Background condition
  & 69.96 & 78.95 & 48.64
  & $-0.12$ & $+0.49$ & $+0.15$ \\
Dominant content
  & 75.34 & 72.29 & 56.20
  & $-0.91$ & $+0.07$ & $+3.44$ \\
Clarity / intelligibility
  & 70.99 & 78.90 & 49.38
  & $+1.63$ & $-1.30$ & $-0.50$ \\
Background sound
  & 67.44 & 65.95 & 48.46
  & $-1.04$ & $+1.17$ & $-0.28$ \\
Descriptive phrases
  & 57.37 & 45.45 & 42.87
  & $+1.78$ & $+5.40$ & $+8.61$ \\
Facets
  & 61.54 & 56.56 & 48.30
  & $+3.01$ & $+4.46$ & $+5.70$ \\
Speaker descriptions
  & 43.51 & 54.12 & 45.66
  & $+1.56$ & $+1.38$ & $-3.57$ \\
Scene hypotheses
  & 70.48 & 57.25 & 50.02
  & $+8.60$ & $+4.15$ & $+2.48$ \\
\midrule
Macro-average
  & \textbf{66.45} & 64.90 & 50.72
  & $+1.35$ & $+2.01$ & $+2.29$ \\
\bottomrule
\end{tabular*}
\end{table}

Table~\ref{tab:app_second_judge_axes} shows that most axes
move by only a few points.
The largest SURE-Tagger shift is scene hypotheses
($+8.60$), while descriptive phrases and facets rise for
all three methods.
Qwen3-Omni-Captioner remains near the half-credit floor on
several axes, consistent with its high descriptive
abstention rate (Appendix~\ref{app:null_rates}).

\begin{table}[t]
\caption{Sample-level agreement between DeepSeek-V4-Pro
and Grok~4.7 on the 648 clips scored by both judges.
The per-sample score is the mean of the ten descriptive
axes.
Grok~$>$ / DS~$>$ count clips whose mean is higher under
Grok~4.7 or DeepSeek-V4-Pro.}
\label{tab:app_second_judge_agree}
\centering
\footnotesize
\setlength{\tabcolsep}{3pt}
\renewcommand{\arraystretch}{1.12}
\begin{tabular*}{\linewidth}{@{\extracolsep{\fill}}l rrr rrrr}
\toprule
Method
  & $n$
  & Pearson
  & Spearman
  & Grok $>$
  & DS $>$
  & Tie
  & Mean $\Delta$ \\
\midrule
SURE-Tagger
  & 648 & 0.929 & 0.924
  & 396 & 250 & 2 & $+1.35$ \\
Qwen3-Omni-Flash
  & 648 & 0.881 & 0.890
  & 388 & 219 & 41 & $+2.01$ \\
Qwen3-Omni-Captioner
  & 648 & 0.764 & 0.753
  & 418 & 219 & 11 & $+2.29$ \\
\bottomrule
\end{tabular*}
\end{table}

Table~\ref{tab:app_second_judge_agree} reports clip-level
agreement on the ten-axis mean.
SURE-Tagger is the most stable across judges
(Pearson $0.929$, Spearman $0.924$).
Qwen3-Omni-Flash remains strongly correlated
($0.881$/$0.890$).
Agreement is weaker for Qwen3-Omni-Captioner
($0.764$/$0.753$), whose outputs are sparser and more often
\texttt{null}.
Grok~4.7 assigns a higher clip-level mean more often than
DeepSeek-V4-Pro for every method, which accounts for the
small positive macro-average shift.

Taken together, the second judge changes absolute
descriptive percentages by about one to two points and
does not change the method ranking on existing-corpus
recordings.
We therefore keep DeepSeek-V4-Pro as the official
descriptive judge in Table~\ref{tab:tagbench} and report
Grok~4.7 only here.

\subsection{Human Layer Preference}
\label{app:human_preference}

We evaluate human layer preferences on 20 existing-corpus
recordings and five representation layers:
L1 lexical content, L2 speaker, L3 temporal structure,
L5 acoustic scenes and events, and D descriptive labels.
This results in
$20 \times 5 = 100$ unique clip--layer items.

To obtain independent judgments while controlling the
annotation cost, we used four annotators.
Each annotator evaluated 50 of the 100 unique items,
resulting in
$4 \times 50 = 200$ judgment records in total.
The assignment was arranged such that each unique item was
evaluated independently by exactly two annotators.
Thus, the study contains 100 unique items rather than 200
unique items, and no item was evaluated by all four
annotators.

Items were shuffled so that the five layers belonging to
the same recording did not appear consecutively.
For each assigned item, an annotator listened to the
recording, inspected only the layer under test, and selected
the most accurate output among SURE-Tagger,
Qwen3-Omni-Flash, Qwen3-Omni-Captioner, and
``cannot decide.''
Each judgment is a single system-level preference rather
than a separate pairwise comparison.

Fields are compared using the ordering
\emph{present and correct} $>$ \emph{absent}
(explicit \texttt{null} or inapplicable) $>$
\emph{incorrect}, following the ordering used in
Equation~\ref{eq:tagbench_score}.
Thus, an incorrect value ranks below leaving the field empty,
whereas a correct additional field ranks above abstention.
The ``cannot decide'' option is reserved for layers that do
not apply to the recording or for cases in which the
annotator cannot reliably rank the systems.

\begin{table}[t]
\caption{Human layer-preference judgments on 100 unique
clip--layer items.
Each item was independently evaluated by two of the four
annotators, yielding 200 judgment records in total.
The $n$ column counts judgment records rather than unique
items.}
\label{tab:app_human_pref}
\centering
\footnotesize
\setlength{\tabcolsep}{4pt}
\renewcommand{\arraystretch}{1.12}
\begin{tabular*}{\linewidth}{@{\extracolsep{\fill}}l r rrrr r}
\toprule
Layer
  & $n$
  & SURE
  & Omni
  & Cap.
  & Undec.
  & SURE \% \\
\midrule
All
  & 200 & 159 & 23 & 1 & 17 & 79.5 \\
L1 lexical content
  & 40 & 24 & 13 & 0 & 3 & 60.0 \\
L2 speaker
  & 40 & 38 & 2 & 0 & 0 & 95.0 \\
L3 temporal structure
  & 40 & 36 & 1 & 0 & 3 & 90.0 \\
L5 scenes and events
  & 40 & 30 & 4 & 0 & 6 & 75.0 \\
D descriptive labels
  & 40 & 31 & 3 & 1 & 5 & 77.5 \\
\bottomrule
\end{tabular*}
\end{table}

Table~\ref{tab:app_human_pref} shows that listeners follow
this ordering.
SURE-Tagger is the clear preference
($159/200=79.5\%$).
The margin is largest on L2 and L3, where it fills speaker
and temporal fields that Captioner leaves \texttt{null} and
that Omni-Flash often leaves incomplete.
Captioner receives a single vote: wholesale abstention is
better than an incorrect fill-in, but it cannot beat a
competitor that already has at least one correct field.
Omni-Flash is the only competitive alternative, mainly on
L1, where a correct transcript with a consistent word count
can outrank SURE-Tagger.
The reverse also holds: an inconsistent derived count
(transcript present but \texttt{word\_count}=0) or an
unverifiable extra noise type on L5 loses to a sparser but
correct Omni-Flash output.
Cannot decide is not a fourth system; it marks inapplicable
layers or indistinguishable items.
The study therefore supports the same field philosophy as
the automatic protocol: write a value when it is known and
correct, leave it empty when it is not, and do not fill the
schema with incorrect values.
\section{Resynthesis-Based Human Preference Evaluation}
\label{app:resynthesis_preference}

\subsection{Resynthesis Pipeline}
\label{app:resynthesis_pipeline}
We selected 30 source recordings and independently extracted semi-structured tags with SURE-Tagger and Qwen-Omni-Captioner (Captioner)~\citep{qwen2025captioner}. Each tag set was converted into a reading script, speaker specifications, and a duration plan. AuK~\citep{ma2026auk} generated a short voice reference for each planned speaker. Base dots.tts~\citep{dots2026tts} synthesized the target speech, after which \texttt{dots.tts.edit}~\citep{wang2026dotsttsedit} adjusted the speaking rate of each segment. Finally, \texttt{saasr-sim-mini} applied reverberation, noise, bandwidth effects, and, where specified by the plan, multi-speaker overlap. The rendered recordings and metadata from each stage were exported. Thus, each source recording yielded two resyntheses that followed the same processing stages but were conditioned on tags from different systems.

\subsection{Listening Comparison}
\label{app:resynthesis_listening}
The listening task presented the source recording together with the two resyntheses, labeled A and B. For each recording, a respondent selected the resynthesis that sounded closer to the source or chose ``undecided.'' We decoded A/B selections using the method mapping stored in the survey records. Two completed questionnaires contained responses for 28 distinct source recordings; 22 recordings appeared in both. We restrict the analysis below to these 22 shared recordings, each with two judgments (44 judgments in total). Recordings appearing in only one questionnaire are excluded from this paired analysis.

\begin{table}[t]
\caption{Human preference for tag-conditioned resynthesis. The response-level percentages use 44 judgments on the 22 shared recordings; the recording-level percentages use the same 22 recordings, each judged twice. Only proportions are shown in the table.}
\label{tab:app_resynthesis_preference}
\centering
\footnotesize
\setlength{\tabcolsep}{8pt}
\renewcommand{\arraystretch}{1.12}
\begin{tabular*}{\linewidth}{@{\extracolsep{\fill}}l r}
\toprule
Outcome & Share (\%) \\
\midrule
\multicolumn{2}{@{}l}{\emph{Preference across individual judgments}} \\
SURE-Tagger resynthesis preferred & 59.1 \\
Captioner resynthesis preferred & 27.3 \\
Undecided & 13.6 \\
\addlinespace[3pt]
\multicolumn{2}{@{}l}{\emph{Agreement across recordings}} \\
Both judgments prefer SURE-Tagger & 36.4 \\
Both judgments prefer Captioner & 0.0 \\
Opposite system preferences & 40.9 \\
One system preference and one undecided & 18.2 \\
Both judgments undecided & 4.5 \\
\bottomrule
\end{tabular*}
\end{table}

\subsection{Results and Interpretation}
\label{app:resynthesis_results}
SURE-Tagger resyntheses received 59.1\% of all judgments, compared with 27.3\% for Captioner; 13.6\% were undecided. Among judgments that selected a system, 68.4\% favored SURE-Tagger. At the recording level, both judgments favored SURE-Tagger for 36.4\% of the shared recordings, while opposite system preferences occurred for 40.9\%. These results indicate an overall perceptual preference in this resynthesis setting, accompanied by substantial variation across judgments. Because the outcome also depends on script construction, voice references, TTS generation and editing, and acoustic rendering, it should be interpreted as an end-to-end listening preference rather than a direct measure of tag accuracy. The number of shared recordings is limited, so we treat this analysis as exploratory and do not claim a statistically established advantage.
\section{Explicit Abstention on Tag-Bench}
\label{app:null_rates}

Section~\ref{sec:experiments} reports Tag-Bench scores in
which an explicit \texttt{null} receives half credit
(Section~\ref{app:tagbench_scoring}).
A method that abstains on every eligible sample therefore
scores $50\%$ on that axis, which is an abstention floor
rather than identification accuracy.
This appendix measures how often each system emits an
explicit \texttt{null} on the existing-corpus subset
(650 recordings; tracks T1--T3) and separates that floor
from identification accuracy.
Structured fields follow the scorer state
\texttt{result\_null}; descriptive fields count a JSON
\texttt{null}, including leaves inherited from a null parent.
An empty array \texttt{[]} or a Boolean \texttt{false} is a
positive assertion and is not counted as abstention.
Missing and runtime-failed states are zero for all three
systems in this snapshot.

Table~\ref{tab:app_null_layers} reports layer-wise
macro-averaged null rates over all 650 recordings.
Qwen-Omni-Captioner is dominated by abstention:
$78.3\%$ of the 29 axes and $89.7\%$ of the 19 structured
axes are explicit \texttt{null}.
On samples that actually enter the score, the structured
null rate remains $87.0\%$.
Its Tag-Bench structured score of $52.70\%$ and overall
score of $51.23\%$ are therefore close to the half-credit
floor, not evidence that about half of the attributes are
identified correctly.

The Captioner's annotations are also limited in coverage
and unstable across related fields.
The free-form \texttt{caption} itself is rarely null
($1.7\%$), but the structured descriptive slots that make
captions comparable are almost empty:
\texttt{facets} $96.3\%$,
\texttt{scene\_hypotheses} $93.7\%$, and
\texttt{descriptive\_phrases} $82.0\%$.
Physical acoustic measurements that never enter the 29-axis
score---SNR, DNSMOS, RT60, and C50---are $100\%$ null, which
is consistent with the captioner control in
Appendix~\ref{app:scene_clustering}.
Even among scored event fields the coverage is inconsistent:
\texttt{external\_noise\_type} is $58.7\%$ null on DESED but
$96.0\%$ null on FSD50K.
A captioner output is therefore a sparse, track-dependent
description rather than a stable attribute inventory.

\begin{table}[t]
\caption{Layer-wise explicit-\texttt{null} rates on the
650 existing-corpus recordings, macro-averaged over axes.
Captioner abstention dominates every scored layer.}
\label{tab:app_null_layers}
\centering
\footnotesize
\setlength{\tabcolsep}{4pt}
\renewcommand{\arraystretch}{1.10}
\begin{tabular*}{\linewidth}{@{\extracolsep{\fill}}l r ccc}
\toprule
Layer & \#
  & SURE-Tagger
  & Captioner
  & Omni-Flash \\
\midrule
L1 lexical content & 6
  & 13.4 & 92.0 & 25.6 \\
L2 speaker & 5
  & 5.5 & 83.3 & 33.2 \\
L3 temporal structure & 4
  & 3.7 & 95.3 & 41.5 \\
L5 acoustic scenes and events & 4
  & 0.0 & 88.7 & 58.0 \\
D descriptive labels & 10
  & 3.9 & 56.7 & 14.6 \\
\midrule
All 29 axes & 29
  & 5.6 & 78.3 & 29.8 \\
\bottomrule
\end{tabular*}
\end{table}

Qwen3-Omni-Flash writes lexical content much more often
(L1 null rate $25.6\%$ overall, $5.1\%$ on eligible L1
samples) but still abstains on acoustic structure:
eligible-sample null rates are $94.9\%$ for
\texttt{nonspeech\_ratio}, $86.9\%$ for
\texttt{speaker\_change\_rate\_per\_min}, and $68.7\%$ for
\texttt{external\_noise\_type}.
The L5 eligible-sample macro-average is $85.2\%$, which
aligns with its L5 score of $55.20\%$ on existing-corpus
recordings.
SURE-Tagger almost always emits a value: $5.6\%$ null over
29 axes, and $2.2\%$ on the 19 structured axes once
unscored T3 speech fields are excluded.
The remaining L1 nulls are largely T3 non-speech clips,
which do not enter the L1 score.
The one material abstention is T1 speaker gender
($133/350=38.0\%$).
SURE-Tagger therefore does not collect the half-credit
floor; the Captioner and Omni-Flash do.

Table~\ref{tab:app_null_vs_score} compares three treatments
of explicit \texttt{null} on the existing-corpus scores from
Table~\ref{tab:tagbench}.
Known-case scores drop null samples and reuse the same
operators and label-first macro-average; axes with no
remaining sample are omitted.
Null-as-zero keeps the original eligible set but scores
\texttt{null} as $0$ rather than $1$.
SURE-Tagger's overall score barely moves
($76.28\rightarrow 76.97$ known-case,
$74.91$ null-as-zero), so the headline percentage is not
an abstention artifact.
The Captioner falls to $12.83\%$ under null-as-zero, and
Omni-Flash to $50.08\%$.
After removing null samples, SURE-Tagger remains highest
overall ($76.97\%$ vs.\ $66.41\%$ and $57.84\%$) and on the
structured axes ($82.77\%$ vs.\ $66.61\%$ and $63.62\%$).
The Captioner known-case omits two structured axes with no
non-null predictions
(\texttt{nonspeech\_ratio} and
\texttt{speaker\_change\_rate\_per\_min}).
A score near $50\%$ is not ``slightly better than chance.''

\begin{table}[t]
\caption{Existing-corpus scores under three treatments of
explicit \texttt{null}.
Known-case drops null samples; null-as-zero keeps them
with score $0$.
Captioner known-case omits two structured axes with no
non-null predictions.}
\label{tab:app_null_vs_score}
\centering
\footnotesize
\setlength{\tabcolsep}{3.5pt}
\renewcommand{\arraystretch}{1.10}
\begin{tabular*}{\linewidth}{@{\extracolsep{\fill}}l ccc}
\toprule
 & SURE-Tagger
  & Captioner
  & Omni-Flash \\
\midrule
Tag-Bench overall (29) & 76.28 & 51.23 & 62.79 \\
Known-case overall & 76.97 & 57.84 & 66.41 \\
Null-as-zero overall & 74.91 & 12.83 & 50.08 \\
\midrule
Tag-Bench structured (19) & 82.16 & 52.70 & 62.74 \\
Known-case structured & 82.77 & 63.62 & 66.61 \\
\midrule
Tag-Bench descriptive (10) & 65.09 & 48.44 & 62.89 \\
Known-case descriptive & 65.96 & 48.03 & 66.04 \\
\midrule
Structured null (eligible) & 2.2 & 87.0 & 30.5 \\
Descriptive null (all 650) & 3.9 & 56.7 & 14.6 \\
\bottomrule
\end{tabular*}
\end{table}

\section{Scene Clustering Details}
\label{app:scene_clustering}

This appendix specifies the attribute encoding, mixed-type
distance, selection rule, cluster inventory, and post-hoc
diagnostics behind the six scene clusters in
Section~\ref{sec:compositional_evaluation}.
It is intended to make the partition auditable without
repeating the corpus-level ASR scores in
Table~\ref{tab:scene_clusters}.
The clusters are frozen from audio-side schema attributes
alone. Dataset identities, transcripts, language text
features, and ASR outputs are not used for assignment, so
the same identifiers can be reused across models.

\subsection{Evaluation Cohort}
The clustering cohort contains 1,000 recordings, 100 from
each of ten sources: AISHELL-1~\citep{bu2017aishell},
AMI~\citep{carletta2006ami}, ChildMandarin~\citep{zhou2024childmandarin},
ContextASR-Bench English and Chinese~\citep{wang2025contextasr},
GigaSpeech2 Thai~\citep{yang2024gigaspeech2},
SeniorTalk~\citep{chen2025seniortalk}, SlideSpeech~\citep{wang2023slidespeech},
SPGISpeech~\citep{oneill2021spgispeech}, and TED-LIUM~\citep{rousseau2012tedlium}.
The language mix is 500 English, 400 Chinese, and 100 Thai
recordings.
English is scored with WER and Chinese/Thai with CER, using
the sidecar metric already chosen for each recording.
ASR evaluation and metric computation use \citet{openbench}.

\subsection{Attribute Vector}
The speech-quality fields include DNSMOS P.835~\citep{reddy2021dnsmos835} scores.
Each recording is encoded as a 23-dimensional mixed-type
vector with a matched missingness mask: 15 numeric fields,
5 binary fields, and 3 set-valued fields.
Missing values are retained as unknown and are never filled
with zeros.
The fields are grouped into four equally weighted axes.

\begin{table}[t]
\caption{Audio-side fields used by $d_{\mathrm{tag}}$.
Source names, language, transcripts, lexical counts, and
ASR errors are excluded.}
\label{tab:app_cluster_fields}
\centering
\footnotesize
\setlength{\tabcolsep}{4pt}
\renewcommand{\arraystretch}{1.08}
\begin{tabularx}{\linewidth}{@{}ll >{\raggedright\arraybackslash}X@{}}
\toprule
Axis & Type & Fields \\
\midrule
Content & binary & music present \\
Content & set
  & speech/music events, external noise type, noise detail \\
Interaction & numeric
  & speaker count, overlap ratio, speaker-change count,
    speaker-change rate \\
Interaction & binary
  & speaker overlap, multi-speaker, speaker change,
    speaker present \\
Quality & numeric
  & non-speech ratio, SNR, DNSMOS (OVRL, BAK, SIG, P.808) \\
Spatial & numeric
  & duration, silence-segment count, total silence, RT60,
    C50 \\
\bottomrule
\end{tabularx}
\end{table}

\subsection{Mixed-type Distance}
Equation~\ref{eq:tag_distance} averages comparable axes.
This section instantiates the per-axis terms.
For a numeric field $f$, a robust center and scale are
fitted on finite values,
\begin{equation}
c_f=\operatorname{median}(\{x_{if}\}),
\qquad
s_f=\operatorname{IQR}(\{x_{if}\}).
\end{equation}
If the IQR is degenerate, the scale falls back to
$1.4826\cdot\mathrm{MAD}$ and then to
$1.4826\cdot\sigma_{\mathrm{pop}}$.
A constant field is dropped from the distance.
Standardized values are $z_{if}=(x_{if}-c_f)/s_f$.
Field-level dissimilarities are
\begin{align}
\delta^{\mathrm{num}}_{ij,f}
&=
\min\bigl(|z_{if}-z_{jf}|/\tau_{\mathrm{num}},\,1\bigr),
\quad \tau_{\mathrm{num}}=3,
\label{eq:app_delta_num}\\
\delta^{\mathrm{bin}}_{ij,f}
&=
\mathbf{1}[x_{if}\neq x_{jf}],
\label{eq:app_delta_bin}\\
\delta^{\mathrm{set}}_{ij,f}
&=
\begin{cases}
0, & A_{if}\cup A_{jf}=\emptyset,\\
1-|A_{if}\cap A_{jf}|/|A_{if}\cup A_{jf}|,
& \text{otherwise.}
\end{cases}
\label{eq:app_delta_set}
\end{align}
Observed dissimilarities are computed only when both
recordings have a value.
For numeric and binary fields, missingness disagreement is
$m_{ij,f}=\mathbf{1}[\mathrm{missing}_{if}\neq\mathrm{missing}_{jf}]$.
A set field contributes $m_{ij,f}=1$ if either side is
missing, including the case where both are missing.
Each axis then mixes observed and missing terms,
\begin{equation}
d_a(i,j)
=(1-\gamma)\,d^{\mathrm{obs}}_a(i,j)
+\gamma\,d^{\mathrm{miss}}_a(i,j),
\qquad \gamma=0.15.
\label{eq:app_axis_distance}
\end{equation}
If an axis has no comparable coordinate, it is omitted from
the mean in Equation~\ref{eq:tag_distance}.
Median/IQR scaling limits the influence of extreme SNR or
duration; the truncation $\tau_{\mathrm{num}}$ limits any
single coordinate; equal axis weights prevent a field-rich
axis from dominating the distance.

\subsection{Search, Selection, and Refinement}
We search $K\in\{6,\ldots,12\}$ with k-medoids~\citep{kaufman1990clustering} (PAM with
$D^2$ initialization~\citep{arthur2007seeding}, seeds $\{0,1,2\}$) and
average-linkage clustering on the same $d_{\mathrm{tag}}$
matrix.
Cluster centers are actual recordings rather than Euclidean
means.
After $D^2$ initialization, each recording is assigned to
its nearest medoid, and each medoid is replaced by the
in-cluster recording that minimizes the sum of within-cluster
distances,
\begin{equation}
m_g
=
\arg\min_{r\in g}
\sum_{i\in g} d_{\mathrm{tag}}(i,r).
\label{eq:app_medoid}
\end{equation}
Assignment and update alternate until the labels stabilize
or an iteration cap is reached.

Candidates are compared in the tag space, not in ASR-error
space.
Let $n_g$ be the size of cluster $g$ and $m_g$ its medoid.
Within-cluster tightness is the size-weighted median
distance to the medoid,
\begin{equation}
C_{\mathrm{tag}}
=
\frac{1}{N}
\sum_{g} n_g\cdot
\operatorname{median}_{i\in g}\,d_{\mathrm{tag}}(i,m_g).
\label{eq:app_ctag}
\end{equation}
Leave-one-source reuse $Q_{\mathrm{tag}}$ is computed by
holding out each source $s$, classifying its recordings to
the remaining medoids, averaging the resulting accuracies
$a_s$, and chance-correcting
\begin{equation}
Q_{\mathrm{tag}}
=
\operatorname{clip}_{[0,1]}
\left(
\frac{\operatorname{mean}_s a_s-1/K}{1-1/K}
\right).
\label{eq:app_qtag}
\end{equation}
The selection cost is
\begin{equation}
\begin{aligned}
J
&=
\lambda_t C_{\mathrm{tag}}
+\lambda_a B_{\mathrm{anchor}}
+\lambda_k\frac{\ln(1+K)}{\ln(1+K_{\max})}
+\lambda_b B_{\mathrm{small}}
+\lambda_r B_{\mathrm{non\text{-}main}},
\end{aligned}
\label{eq:app_j}
\end{equation}
with $\lambda_t=1$, $\lambda_a=\lambda_b=\lambda_r=0.8$,
$\lambda_k=0.15$, and $K_{\max}=12$.
$B_{\mathrm{anchor}}$ penalizes missing tag anchors or
insufficient $Q_{\mathrm{tag}}$;
$B_{\mathrm{small}}$ penalizes clusters smaller than
$n_{\min}=30$;
$B_{\mathrm{non\text{-}main}}$ is the fraction of
unassigned recordings.
A candidate is discarded as source-driven if any cluster
with $n_g\ge n_{\min}$ has a single-source share of at
least $0.90$.

The selected k-medoids solution had $K=7$.
A subsequent tag-only merge step combined two clusters
whose attribute profiles were not distinguishable, while
retaining a genuine 17-recording tail cluster, and the
labels were canonicalized to C00--C05.
No ASR error enters this refinement.
The frozen partition has $K=6$,
$C_{\mathrm{tag}}=0.071$, $Q_{\mathrm{tag}}=0.606$,
$J=0.195$, $B_{\mathrm{anchor}}=B_{\mathrm{non\text{-}main}}=0$,
and is not source-driven.
A mutual-nearest-neighbor cut diagnostic, not used in $J$,
is $C_{\mathrm{cut}}=0.031$: only about $3.1\%$ of local
similar edges cross cluster boundaries.
New recordings, including the fine-tuning candidate pool
in Section~\ref{sec:targeted_selection}, are assigned to
the nearest frozen medoid
$\hat z(x)=\arg\min_g d_{\mathrm{tag}}(x,m_g)$.

\subsection{Cluster Inventory and Scoring Coverage}
Table~\ref{tab:app_cluster_inventory} reports cluster sizes
and median acoustic evidence.
The names match Table~\ref{tab:scene_clusters} and are
observational summaries, not location labels.
C00--C03 cover $94.5\%$ of the cohort and correspond to
ordinary single-speaker speech at different duration, SNR,
and reverberation regimes.
C04 and C05 are small boundary clusters ($1.7\%$ and
$3.8\%$): very short speech-sparse fragments and short
low-SNR clips with music.
They are retained because their tag profiles are distinct,
but they should not be read as stable room types.

\begin{table}[t]
\caption{Frozen scene clusters ($N=1000$).
Duration, SNR, and RT60 are cluster medians.}
\label{tab:app_cluster_inventory}
\centering
\footnotesize
\setlength{\tabcolsep}{3.5pt}
\renewcommand{\arraystretch}{1.08}
\begin{tabular*}{\linewidth}{@{\extracolsep{\fill}}lrrrrrl}
\toprule
ID & $n$ & \% & Dur.\ (s) & SNR (dB) & RT60 (s) & Profile \\
\midrule
C00 & 341 & 34.1 & 6.62 & 35.6 & 0.27 & Mid-length, dry speech \\
C01 & 122 & 12.2 & 1.25 & 17.3 & 0.37 & Short, low SNR \\
C02 & 286 & 28.6 & 3.04 & 33.8 & 0.30 & Short, moderate reverb \\
C03 & 196 & 19.6 & 45.56 & 66.0 & 0.20 & Long, clean speech \\
C04 & 17 & 1.7 & 0.42 & 36.1 & 0.48 & Very short, speech-sparse \\
C05 & 38 & 3.8 & 0.70 & 12.6 & 0.44 & Short, noisy with music \\
\bottomrule
\end{tabular*}
\end{table}

Source membership is mixed rather than one-to-one.
Table~\ref{tab:app_cluster_sources} lists the datasets
present in each cluster.
C03 contains only the two ContextASR-Bench language splits,
which are counted as distinct sources, so it is not flagged
as source-driven even though its acoustic profile is highly
distinctive.

\begin{table}[t]
\caption{Data sources present in each frozen scene cluster.}
\label{tab:app_cluster_sources}
\centering
\footnotesize
\setlength{\tabcolsep}{4pt}
\renewcommand{\arraystretch}{1.08}
\begin{tabularx}{\linewidth}{@{}l >{\raggedright\arraybackslash}X@{}}
\toprule
ID & Sources \\
\midrule
C00 & AISHELL-1, AMI, ChildMandarin, ContextASR-Bench English,
      GigaSpeech2 Thai, SeniorTalk, SlideSpeech, SPGISpeech,
      TED-LIUM \\
C01 & AMI, ChildMandarin, GigaSpeech2 Thai, SeniorTalk,
      SlideSpeech \\
C02 & AISHELL-1, AMI, ChildMandarin, ContextASR-Bench English,
      GigaSpeech2 Thai, SeniorTalk, SlideSpeech, SPGISpeech,
      TED-LIUM \\
C03 & ContextASR-Bench Chinese, ContextASR-Bench English \\
C04 & AMI, SeniorTalk \\
C05 & AMI, GigaSpeech2 Thai \\
\bottomrule
\end{tabularx}
\end{table}

Table~\ref{tab:app_cluster_coverage} reports per-model
scoring coverage.
Parakeet-TDT (using the token-and-duration transducer
architecture~\citep{xu2023tdt}), MOSS-Transcribe~\citep{yu2026moss},
and Granite Speech~\citep{granite2026speech} score the
500 English recordings only.
Whisper large-v3 turbo~\citep{radford2023whisper,openai2024whisperturbo}
and SenseVoiceSmall~\citep{an2024funaudiollm} score English and
Chinese ($n=900$) and omit the 100 Thai recordings.
Qwen3-ASR-1.7B~\citep{shi2026qwen3asr} and
FunASR-Flash~\citep{funasr2026flash} cover all 1,000 recordings,
including later-completed Thai CER for Qwen.
Coverage is therefore a property of language support, not of
the frozen partition.
Cross-model comparisons in
Table~\ref{tab:scene_clusters} should be read together with
these denominators, especially on C01 and C05, where Thai
recordings are concentrated.
Insertions enter the numerator without increasing the
reference denominator, so a reported rate may exceed
$100\%$.

\begin{table}[t]
\caption{Valid scored recordings per cluster.
The three English-only models share one coverage pattern;
Whisper and SenseVoice share another.}
\label{tab:app_cluster_coverage}
\centering
\footnotesize
\setlength{\tabcolsep}{4pt}
\renewcommand{\arraystretch}{1.08}
\begin{tabularx}{\linewidth}{@{}>{\raggedright\arraybackslash}X rrrr@{}}
\toprule
Cluster & $n$
  & English-only
  & Whisper / SenseVoice
  & Qwen / FunASR \\
\midrule
C00 & 341 & 235 & 321 & 341 \\
C01 & 122 & 33 & 62 & 122 \\
C02 & 286 & 99 & 283 & 286 \\
C03 & 196 & 96 & 196 & 196 \\
C04 & 17 & 16 & 17 & 17 \\
C05 & 38 & 21 & 21 & 38 \\
\midrule
Total & 1000 & 500 & 900 & 1000 \\
\bottomrule
\end{tabularx}
\end{table}

\subsection{Post-hoc Error Homogeneity}
After the partition is frozen, we test whether recordings in
the same cluster also have more similar ASR errors.
This check is diagnostic: it does not revise
$d_{\mathrm{tag}}$, $K$, or $\mathrm{cluster\_id}$.
For a model $m$, let $\mathcal{V}$ be the recordings with a
valid per-sample error $e_i$ (WER or CER), $N=|\mathcal{V}|$,
and $\mathcal{V}_g$ the valid subset of cluster $g$.
Then
\begin{equation}
\begin{aligned}
\mu
&=
\frac{1}{N}\sum_{i\in\mathcal{V}} e_i,
\qquad
\mu_g
=
\frac{1}{n_g}\sum_{i\in\mathcal{V}_g} e_i,\\
\eta^2
&=
1-
\frac{\sum_g\sum_{i\in\mathcal{V}_g}(e_i-\mu_g)^2}
{\sum_{i\in\mathcal{V}}(e_i-\mu)^2}.
\end{aligned}
\label{eq:app_eta2}
\end{equation}
$\eta^2$ is the fraction of error variance associated with
between-cluster mean differences.
It is not a classification accuracy and not a proof that
the clusters are true acoustic scenes.
Because maximizing $\eta^2$ favors larger $K$, we also
report
\begin{equation}
S_{\eta K}
=
\eta^2
-
\lambda\log_2\bigl(\max(K,K_{\mathrm{ref}})/K_{\mathrm{ref}}\bigr),
\label{eq:app_setak}
\end{equation}
with $\lambda=0.3$ and $K_{\mathrm{ref}}=6$.
The complexity cost is zero for $K\le 6$ and $0.221$ for
$K=10$.

These per-sample statistics are not interchangeable with
Table~\ref{tab:scene_clusters}.
The main table pools edits over reference units, whereas
$\eta^2$ treats each recording equally.
The distinction is sharpest on C05 for Qwen3-ASR-1.7B: the
per-sample median remains $100\%$ because 21 very short
references saturate, while the micro-averaged CER falls to
$17.26\%$ after the 17 Thai recordings are included.

Table~\ref{tab:app_homogeneity} compares four groupings on
the same valid errors: a size-matched random shuffle
(80 repetitions), the frozen tag clusters, grouping by
source, and a diagnostic 6-bin oracle obtained by sorting
errors and cutting them into contiguous segments.
The oracle is an upper bound for error-based binning, not a
scene definition that can be reused across models.
For Qwen, the tag partition yields $\eta^2=0.348$ versus
$0.005$ for the random baseline.
Source grouping attains a higher raw $\eta^2$ with ten
groups, but its complexity-adjusted score is
$S_{\eta K}=0.366$, only slightly above the 6-cluster tag
score of $0.348$.
Whisper and FunASR still favor the tag partition over
source grouping on $S_{\eta K}$; SenseVoice is only weakly
structured by these clusters.
The same identifiers therefore expose model-specific
difficulty patterns; they are not a universal difficulty
ranking and do not replace a model's overall score.

The headline Qwen value is sensitive to the two tail
clusters.
Excluding C05 leaves $\eta^2=0.294$ ($n=962$);
excluding C04 and C05 leaves $\eta^2=0.144$ ($n=945$).
C04/C05 should be reported with their sizes, or omitted
from headline ASR averages, rather than treated as ordinary
scene bins.

\begin{table}[t]
\caption{Post-hoc error homogeneity.
$S_{\eta K}=\eta^2$ for every $K=6$ grouping.
Source grouping uses $K=10$ when Thai scores are present
and $K=9$ otherwise.
The 6-bin oracle sorts errors and is not a scene partition.}
\label{tab:app_homogeneity}
\centering
\footnotesize
\setlength{\tabcolsep}{3.2pt}
\renewcommand{\arraystretch}{1.08}
\begin{tabularx}{\linewidth}{@{}>{\raggedright\arraybackslash}X rrrrrr@{}}
\toprule
Model & $n$
  & Rand.\ $\eta^2$
  & Tag $\eta^2$
  & Src $\eta^2$
  & Src $S_{\eta K}$
  & Oracle $\eta^2$ \\
\midrule
Qwen3-ASR-1.7B
  & 1000 & 0.005 & 0.348 & 0.587 & 0.366 & 0.980 \\
Whisper large-v3 turbo
  & 900 & 0.006 & 0.366 & 0.392 & 0.216 & 0.963 \\
SenseVoiceSmall
  & 900 & 0.006 & 0.060 & 0.178 & 0.003 & 0.964 \\
FunASR-Flash
  & 1000 & 0.005 & 0.266 & 0.462 & 0.241 & 0.978 \\
\bottomrule
\end{tabularx}
\end{table}

\subsection{Caption-derived Control}
We repeat the same tags-only procedure on caption-derived
attributes for the same 1,000 recordings.
The captioner preserves speaker-interaction and
event/noise mentions, but SNR, DNSMOS, RT60, and C50 are
entirely null, and duration is missing for $94.6\%$ of
recordings.
The quality and spatial axes therefore contribute almost
no distance.
The frozen captioner partition is still k-medoids with
$K=6$ ($J=0.140$, $Q_{\mathrm{tag}}=0.605$) and is not
source-driven, yet its adjusted Rand index with the
SURE-Tagger partition is only $0.036$.
On Qwen scores, the captioner clusters yield
$\eta^2=S_{\eta K}=0.022$, comparable to a random shuffle
($0.005$) and far below the SURE-Tagger value of $0.348$;
the FunASR captioner score is $S_{\eta K}=0.034$.
A compact captioner partition can still be formed, but it
does not explain ASR errors and is not a substitute for
the SURE-Tagger clusters used in
Section~\ref{sec:compositional_evaluation}.

%

\section{Supplementary Construction Details of Tag-Bench}
\label{app:tag_bench_details}

The main text presents the overall motivation and high-level construction
pipeline of Tag-Bench. This appendix reports the operational definitions,
agentic orchestration process, rendering parameters, sampling rules, and
reference-building details required to reproduce the benchmark.

\subsection{Speaker-interaction pattern catalog}

Table~\ref{tab:interaction_patterns} lists the 11 valid interaction patterns
and their frequencies in the final 500 synthetic scenes.

\begin{table}[t]
\centering
\small
\caption{Speaker-interaction patterns in the synthetic subset.}
\label{tab:interaction_patterns}
\begin{tabularx}{\linewidth}{@{}c >{\raggedright\arraybackslash}X c@{}}
\toprule
ID & Operational pattern & Scenes \\
\midrule
I01 & Single-speaker monologue & 75 \\
I02 & Two-speaker long-turn alternation without overlap & 43 \\
I03 & Two-speaker rapid turn alternation without overlap & 42 \\
I04 & Dominant-speaker speech; the second speaker appears only in overlap & 43 \\
I05 & Two-speaker sparse turn-taking with overlap & 42 \\
I06 & Two-speaker frequent turn-taking with overlap & 43 \\
I07 & Multi-speaker sequential long-turn discussion & 42 \\
I08 & Multi-speaker frequent turn-taking without overlap & 43 \\
I09 & Dominant-speaker speech; all other speakers appear only in overlap & 42 \\
I10 & Multi-speaker sparse turn-taking with overlap & 43 \\
I11 & Multi-speaker frequent turn-taking with competitive overlap & 42 \\
\midrule
& \textbf{Total} & \textbf{500} \\
\bottomrule
\end{tabularx}
\end{table}

The catalog enforces logical constraints before rendering. A single-speaker
scene cannot contain speaker changes or overlapping speakers. Overlap requires
at least two audible speaker stems. A multi-speaker specification with neither
speaker changes nor overlap is invalid because the additional speakers would
never become audible. In I04 and I09, the non-dominant speakers do not receive
independent non-overlapping turns; their speech occurs only while the dominant
speaker is active. I05, I06, I10, and I11 contain both sequential changes and
overlap, with sparse or frequent switching as specified by the pattern.

After these interaction constraints are resolved, the 11 valid patterns are
crossed with three speaking-rate levels, three SNR levels, two music states,
and two reverberation levels. The controlled base space is therefore

\begin{equation}
    \mathcal{C}_{\mathrm{base}}
    = \mathcal{I}_{11}
    \times \mathcal{R}_{3}
    \times \mathcal{S}_{3}
    \times \mathcal{M}_{2}
    \times \mathcal{V}_{2},
    \qquad
    \lvert\mathcal{C}_{\mathrm{base}}\rvert
    = 11 \times 3 \times 3 \times 2 \times 2
    = 396,
\end{equation}

where $\mathcal{I}$ denotes the interaction pattern, $\mathcal{R}$ speaking
rate, $\mathcal{S}$ SNR level, $\mathcal{M}$ music absence or presence, and
$\mathcal{V}$ reverberation level. Thus, 396 counts the Cartesian product of
the already validated interaction patterns and the remaining controlled axes;
it is not the unconstrained product of all raw schema attributes. Alternative
scripts and subsequent lexical and multilingual extensions are introduced
after this base grid is formed.

\subsection{Acoustic rendering parameters}

\subsubsection{SNR and reverberation levels}

The SNR catalog uses the ranges and generation targets in
Table~\ref{tab:snr_levels}. The renderer targets the center value, whereas the
strict reference stores the SNR achieved in the final mixture.

\begin{table}[t]
    \centering
    \small
    \caption{SNR catalog levels.}
    \label{tab:snr_levels}
    \begin{tabularx}{\linewidth}{@{}>{\raggedright\arraybackslash}X cc@{}}
        \toprule
        Level & Accepted range & Generation target \\
        \midrule
        Low SNR & 3--8 dB & 5 dB \\
        Medium SNR & 12--18 dB & 15 dB \\
        High SNR & 24--30 dB & 27 dB \\
        \bottomrule
    \end{tabularx}
\end{table}

The two reverberation levels are jointly defined by RT60 and C50. Catalog
targets guide RIR selection, but the strict reference values are the measured
RT60 and C50 of the RIR actually selected for a scene.

\begin{table}[t]
    \centering
    \small
    \caption{Reverberation catalog levels.}
    \label{tab:reverb_levels}
    \begin{tabularx}{\linewidth}{@{}>{\raggedright\arraybackslash}X cccc@{}}
        \toprule
        Level & RT60 range & Target RT60 & C50 range & Target C50 \\
        \midrule
        Lower reverberation & 0.15--0.40 s & 0.28 s & 8--18 dB & 12 dB \\
        Higher reverberation & 0.80--1.40 s & 1.10 s & $-2$--5 dB & 2 dB \\
        \bottomrule
    \end{tabularx}
\end{table}

\subsubsection{Background and music scaling}

Music is not added using a fixed final music-to-speech gain. Each music source
is converted to mono at 24~kHz, DC-removed, and normalized to RMS 0.3270884.
The internal background mixture is then

\begin{equation}
    b_{\mathrm{raw}}(t)
    = 0.55 b_{\mathrm{amb}}(t)
    + 0.75 b_{\mathrm{event}}(t)
    + 0.85 b_{\mathrm{music}}(t),
\end{equation}

where the components are ambient noise, discrete background events, and real
music. The complete background is normalized to RMS 0.10 and attenuated again
if its peak exceeds 0.9. Finally, the entire background track is rescaled
against reverberant speech over speech-active intervals to reach the requested
5, 15, or 27~dB SNR. Therefore, 0.85 is an internal music weight rather than
the final music-to-speech gain.

Let $s_{\mathrm{rev}}(t)$ denote the summed reverberant speech and
$b_{\mathrm{final}}(t)$ the final composite background after rescaling. The
achieved SNR is

\begin{equation}
    \mathrm{SNR}_{\mathrm{achieved}}
    = 10\log_{10}
    \frac{\sum_{t \in \Omega_{\mathrm{speech}}} s_{\mathrm{rev}}^2(t)}
         {\sum_{t \in \Omega_{\mathrm{speech}}} b_{\mathrm{final}}^2(t)},
\end{equation}

where $\Omega_{\mathrm{speech}}$ denotes speech-active intervals. For scenes
with music, the denominator includes music together with ambient noise and
discrete events.

\subsection{Accounting for lexical and multilingual variants}

The reported lexical statistics contain overlap and should not be interpreted
as 104 independent additions. The initial realized pool contained 468 scenes.
Of these, 48 scripts were rewritten to introduce repetition and fillers: 24
Chinese code-switching scripts and 24 English scripts without code-switching.
The Chinese code-switching scenes are therefore a subset of the 48
repetition-and-filler scenes. A further 32 scenes were added, producing the
final total of 500.

\begin{table}[t]
    \centering
    \small
    \caption{Mutually exclusive accounting of the lexical modifications.}
    \label{tab:lexical_expansion}
    \begin{tabularx}{\linewidth}{@{}>{\raggedright\arraybackslash}X cc@{}}
        \toprule
        Scene type & Scenes & Operation \\
        \midrule
        Chinese code-switching with repetition and fillers & 24 & Rewrite \\
        English repetition and fillers without code-switching & 24 & Rewrite \\
        Monolingual scenes in eight additional languages & 16 & Add \\
        Multilingual dialogues & 16 & Add \\
        \midrule
        Unique scenes modified or added & 80 & 48 rewritten; 32 added \\
        \bottomrule
    \end{tabularx}
\end{table}

The eight additional languages are Arabic (\texttt{ar}), Spanish
(\texttt{es}), French (\texttt{fr}), Hindi (\texttt{hi}), Japanese
(\texttt{ja}), Korean (\texttt{ko}), Russian (\texttt{ru}), and Swahili
(\texttt{sw}), with two monolingual scenes per language.

The 16 multilingual dialogues are divided into two groups. Eight use a fixed
language for each speaker: \texttt{zh-en}, \texttt{zh-ja}, \texttt{zh-ko},
\texttt{en-es}, \texttt{en-fr}, \texttt{en-hi}, \texttt{ar-en-sw}, and
\texttt{ru-zh-fr}. In the other eight, the same speaker may switch languages
across turns: \texttt{zh-en}, \texttt{zh-ja}, \texttt{zh-ko},
\texttt{en-es}, \texttt{en-fr}, \texttt{en-hi}, \texttt{ar-en}, and
\texttt{ru-zh}.

\subsection{Agentic synthesis orchestration}

We use \emph{agentic synthesis} to describe specification-driven orchestration
rather than unconstrained waveform generation. The agentic planner operates
above a set of fixed generation and signal-processing tools. It interprets the
schema-derived scene specification, checks cross-field constraints, decomposes
each scene into turn-level jobs, selects the applicable tool path, records
intermediate state, and applies condition-dependent branches or fallbacks.
Waveform transformations and reference calculations remain deterministic and
auditable; model predictions are not used as synthetic ground truth.

\subsubsection{Versioned planning and constraint propagation}

The planner maintains a versioned scene catalog so that design changes can be
introduced without silently changing the controlled acoustic grid. Version~1
created 396 valid L2--L5 configurations and 72 alternative lexical
realizations. Version~2 naturalized selected scripts, added planned turn
timestamps, and selected environment-compatible RIRs. Version~3 introduced
Chinese--English coverage and balanced reference-voice assignment while
preserving the inherited acoustic conditions. Version~4 added code-switching,
eight additional languages, and multilingual interactions, producing the final
500 scene specifications.

Voice assignment is also handled as a constrained allocation problem. A voice
cannot be reused by different speakers within the same session. Across
sessions, candidate voices are selected using usage counts and recent
assignments to avoid repeatedly choosing the same prompts. English voices are
drawn from AMI~\citep{carletta2006ami}; Chinese voices are drawn from
AliMeeting~\citep{yu2021alimeeting} and AISHELL-3~\citep{shi2020aishell3};
the additional languages use language-matched FLEURS~\citep{conneau2022fleurs} prompts.

\subsubsection{Compilation into a recoverable task graph}

Speech generation uses dots.tts~\citep{dots2026tts}, and
rate editing uses dots.tts.edit~\citep{wang2026dotsttsedit}.
The final scene JSONL is compiled into a six-stage dependency graph:

\begin{equation}
\texttt{prepare} \rightarrow \texttt{base} \rightarrow \texttt{edit}
\rightarrow \texttt{derive} \rightarrow \texttt{render} \rightarrow
\texttt{qc}.
\end{equation}

Each stage writes explicit artifacts and can be resumed independently. The
planner therefore reasons over scene-level requirements, while the executor
operates on cached jobs and versioned manifests. The complete compilation
produces 2,071 base TTS sources, 1,363 rate-edit sources, 2,089 derived sources,
and 500 session render specifications. Table~\ref{tab:agentic_stages}
summarizes the decisions and outputs of each stage.

\begin{table}[t]
    \centering
    \scriptsize
    \caption{Agent-orchestrated synthesis stages and deterministic executors.}
    \label{tab:agentic_stages}
    \begin{tabularx}{\linewidth}{@{}>{\raggedright\arraybackslash}p{0.12\linewidth}>{\raggedright\arraybackslash}p{0.34\linewidth}>{\raggedright\arraybackslash}X@{}}
        \toprule
        Stage & Planner or controller decision & Deterministic execution and recorded output \\
        \midrule
        Prepare & Parse every scene and turn; validate required fields; route
        jobs by language, voice, rate, and rendering condition & Deduplicate
        base sources using the tuple of voice, language, and text; derive a
        fixed seed from the job key; emit dialogue JSON, edit jobs, derived jobs,
        and a unified render matrix \\
        Base & Select the language-matched voice prompt and decide whether long
        text must be chunked & Generate speech using \texttt{dots.tts-base};
        retain 24-kHz working audio and 48-kHz native audio; trim abnormal edge
        silence and concatenate chunks with controlled pauses \\
        Edit & Route slow and fast items to rate editing; allow normal-rate
        items to bypass this stage & Request factors 0.67 and 1.50 using
        \texttt{dots.tts.edit}; record checkpoint use or
        \texttt{fallback\_time\_stretch} when editing is unavailable \\
        Derive & Compare the requested rate with the realized active-speech
        span and decide whether further correction is required & Trim edge
        silence, compress abnormally long internal pauses, apply FFmpeg
        \texttt{atempo}, calibrate active RMS, and enforce a 0.98 peak ceiling \\
        Render & Resolve the interaction template, RIR path, background assets,
        and target SNR for the session & Schedule turns; construct per-speaker
        dry stems; convolve them with the selected RIR; sum wet stems; add one
        continuous scene-level background; rescale for SNR; export mixture,
        stems, RTTM, and session manifest \\
        QC & Compare rendered evidence with structural and acoustic constraints;
        separate hard failures from dimension-specific warnings & Check WAV
        integrity, channel count, sample rate, speaker count, overlap, achieved
        SNR, RIR measurements, and source-rate proxies; emit a machine-readable
        QC report \\
        \bottomrule
    \end{tabularx}
\end{table}

The orchestration contains explicit conditional branches. Normal-rate items
skip rate editing, whereas slow and fast items request edit factors and then
undergo deterministic rate calibration. Of the 1,363 edited sources, 1,314 use
the edit checkpoint and 49 record a time-stretch fallback. For room rendering,
the planner selects a measured RIR when its environment and measured RT60/C50
match the scene within tolerance; otherwise it routes the item to a synthetic
RIR. The final set contains 305 measured-RIR scenes and 195 synthetic-RIR
scenes. All speakers within a session share a room-consistent RIR choice.

\subsubsection{Turn scheduling and session rendering}

The planner stores turn-level start and end times as an initial schedule, but
does not treat planned times as audio alignment. After TTS and rate calibration,
the renderer reschedules the realized utterances according to the interaction
template. Each speaker first receives an independent dry stem, followed by a
room-processed wet stem. Only after all wet stems are combined is the continuous
background track added and scaled to the requested SNR. This ordering prevents
background noise from switching on and off at speaker boundaries and preserves
a shared acoustic scene across the complete session.

The background asset is assembled at scene level from the environment profile,
discrete events, and optional music. The renderer then exports the final
mixture together with component stems, RTTM, the realized turn sequence, and a
session manifest. These intermediate artifacts provide the evidence required
for post-render checks and reference construction.

\subsubsection{Post-render verification, reference recalculation, and execution status}

The final waveform is frozen before QC. After this point, verification is a
read-only operation: it does not trigger re-synthesis, time stretching,
re-mixing, gain adjustment, or RIR replacement. This separation avoids making
the evaluated audio less natural merely to force it to match a planned catalog
value. If a requested condition and its final realization differ, we update the
corresponding reference annotation from post-render evidence rather than modify
the audio to fit the original specification. The requested value is retained
as provenance, whereas the remeasured value becomes the ground truth used for
evaluation.

SNR illustrates this procedure. The values 5, 15, and 27~dB guide background
scaling during rendering, but they are not copied directly into the reference
annotation. Once the mixture has been finalized, we recompute SNR over
speech-active regions from the post-RIR speech stems and the finalized
background stem:
\begin{equation}
    \mathrm{SNR}_{\mathrm{GT}}
    =10\log_{10}
    \frac{\sum_{t\in\mathcal{A}} s_{\mathrm{postRIR}}^{2}(t)}
         {\sum_{t\in\mathcal{A}} b_{\mathrm{final}}^{2}(t)},
\end{equation}
where $\mathcal{A}$ is the union of realized speech-active intervals. The
measured value, rather than the intended SNR target, is written to the strict
ground truth. The same principle is applied to other directly recoverable
attributes: duration and format are read from the final WAV header; overlap and
speaker changes are recomputed from the realized schedule and stem activity;
and RT60/C50 use measurements of the RIR actually selected for rendering.

The QC stage additionally checks decodability, mono format, sample rate,
non-empty waveforms, speaker count, overlap, RIR consistency, and a source-rate
proxy. All 500 scenes were rendered, totaling 3.7784 hours, with zero hard
failures. The report retains 324 warnings: 304 source-rate proxy mismatches and
20 deviations between the measured C50 of a selected RIR and its target catalog
level. Thus, \emph{QC passed} indicates that no structural hard failure
occurred, not that every planned value was realized exactly. A warning records
the target--realization discrepancy, while any reference field supported by
direct post-render evidence uses its recalculated value. Fields that cannot be
established reliably are excluded for the corresponding evaluation or set to
\texttt{null}; SURE-Tagger predictions are never used to repair the reference.

\subsection{Rendering records and reference construction}

For each synthetic item, the renderer retains the frozen script, speaker and
voice assignments, isolated speech stems, realized turn schedule, RIR
identifier, background-source identifiers, mixing coefficients, random seed,
requested conditions, achieved measurements, and QC status. These records are
used to construct field-level references without additional inference, as
summarized in Table~\ref{tab:synthetic_reference}.

\begin{table}[t]
    \centering
    \small
    \caption{Evidence used to establish synthetic reference annotations.}
    \label{tab:synthetic_reference}
    \begin{tabularx}{\linewidth}{@{}>{\raggedright\arraybackslash}p{0.25\linewidth}>{\raggedright\arraybackslash}X@{}}
        \toprule
        Attribute & Reference evidence \\
        \midrule
        Duration, sample rate, channels & Final WAV header \\
        Transcript and text statistics & Frozen script \\
        Language spans & Turn-level language and \texttt{language\_spans} in the
        generation metadata \\
        Speaker count and overlap & Realized session manifest and stem activity \\
        Speaker-change count & Actual rendered turn sequence \\
        Speaking-rate evidence & Linguistic-unit counts and active-speech spans
        of the derived sources \\
        Ground-truth SNR & SNR recomputed from post-RIR speech stems and the
        finalized background stem \\
        Frame-level SNR & Framewise calculation from post-RIR speech and noise stems \\
        RT60 and C50 & Measurements of the RIR selected for the rendered session \\
        Music presence & Music-source record and inclusion in the scene-level
        background track \\
        Unsupported fields & \texttt{null} for DNSMOS, far-field status, speaker
        profiles, and descriptive attributes when no direct evidence exists \\
        \bottomrule
    \end{tabularx}
\end{table}

The overlap ratio is computed from the realized speaker timeline as

\begin{equation}
    R_{\mathrm{overlap}}
    = \frac{\left|\{t:\sum_i \mathbf{1}_i(t) \geq 2\}\right|}
           {\left|\{t:\sum_i \mathbf{1}_i(t) \geq 1\}\right|},
\end{equation}

where $\mathbf{1}_i(t)$ indicates whether speaker $i$ is active at time $t$.
This makes the reference depend on the realized timeline rather than only on a
binary overlap request. Code-switching and multilingual items cannot always be
represented by one sample-level language code. Accordingly, 40 synthetic items
use \texttt{language\_content.language=null}, while their actual language lists
and turn-level spans remain available in the generation metadata.

\subsection{Existing-corpus sampling details}

Table~\ref{tab:existing_corpora} reports the exact sampling procedure for the
650 existing-corpus items, totaling 193.05 minutes (approximately 3.22 hours).
The source corpora are AMI~\citep{carletta2006ami},
FLEURS~\citep{conneau2022fleurs}, LibriSpeech~\citep{panayotov2015librispeech},
DESED~\citep{turpault2019desed}, and FSD50K~\citep{fonseca2020fsd50k}.
This subset is not uniformly real speech because the selected DESED split
contains synthetic soundscapes.

\begin{table}[t]
    \centering
    \scriptsize
    \caption{Existing-corpus composition and sampling rules.}
    \label{tab:existing_corpora}
    \begin{tabularx}{\linewidth}{@{}>{\raggedright\arraybackslash}p{0.13\linewidth}c >{\raggedright\arraybackslash}p{0.18\linewidth}>{\raggedright\arraybackslash}X@{}}
        \toprule
        Corpus & Items & Split or source & Sampling rule \\
        \midrule
        AMI & 350 & Mix-Headset; no official split filtering & All 170 meetings
        in the local utterance manifests; two clips per meeting, with ten
        meetings contributing a third clip \\
        FLEURS & 100 & \texttt{test} & Ten items from each locale:
        \texttt{ar\_eg}, \texttt{cmn\_hans\_cn}, \texttt{en\_us},
        \texttt{es\_419}, \texttt{fr\_fr}, \texttt{hi\_in},
        \texttt{ja\_jp}, \texttt{ko\_kr}, \texttt{ru\_ru}, and
        \texttt{sw\_ke} \\
        LibriSpeech & 50 & \texttt{test.clean} & 32 utterances from speaker
        1089 and 18 from speaker 1188 \\
        DESED & 75 & \texttt{synthetic21\_}\newline\texttt{train} & Strongly annotated
        10.00-second soundscapes from the 16-kHz synthetic soundscape subset \\
        FSD50K & 75 & \texttt{dev}: 38; \texttt{eval}: 37 & Clip-level verified
        multi-label sound-event annotations \\
        \midrule
        \textbf{Total} & \textbf{650} & -- & \textbf{193.05 min} \\
        \bottomrule
    \end{tabularx}
\end{table}

Native corpus annotations are retained at their original granularity. For
example, clip-level FSD50K labels are not converted into event timestamps, and
an unlisted event class is not treated as confirmed absence. Missing attributes
may be proposed by Qwen3.5-Omni-Plus~\citep{qwen2026omniplus}, but these proposals enter the reference
only after manual review. Exact physical quantities, including true source SNR,
source distance, or room impulse response, are not assigned from the multimodal
model alone. Attributes without sufficient evidence remain \texttt{null}.

The corpus selections also impose two reporting constraints. First, the AMI
subset draws from all 170 local meeting manifests rather than an official
train/development/test partition; source meeting IDs must therefore be retained
for leakage auditing. Second, the current LibriSpeech subset contains only two
speakers and should be interpreted as clean read-speech coverage rather than
broad speaker diversity.

\section{Training details and Whisper-small Adaptation under the Same Budget}
\label{app:whisper_selection}

\begin{table}[t]
\caption{Fine-tuning settings for the one-hour adaptation of
Qwen3-ASR-1.7B (Section~\ref{sec:targeted_selection}) and
Whisper-small.}
\label{tab:app_ft_hparams}
\centering
\footnotesize
\setlength{\tabcolsep}{3pt}
\renewcommand{\arraystretch}{1.12}
\begin{tabular*}{\linewidth}{@{\extracolsep{\fill}}lcc}
\toprule
Setting
  & Qwen3-ASR-1.7B
  & Whisper-small \\
\midrule
Recipe
  & Official Qwen3-ASR SFT
  & HF Whisper SFT \\
Parameters
  & Full model
  & Full model \\
GPUs
  & $2\times$ RTX 4090
  & $1\times$ RTX 4090 \\
Precision
  & bf16
  & fp16 \\
Batch (device / acc.\ / global)
  & $1$ / $128$ / $256$
  & $16$ / $1$ / $16$ \\
Learning rate
  & $2\times10^{-5}$
  & $1\times10^{-5}$ \\
Scheduler
  & linear, 2\% warmup
  & linear, 10\% warmup \\
Optimizer
  & AdamW, wd$=0$, clip $1.0$
  & AdamW, wd$=0$, clip $1.0$ \\
Grad.\ checkpointing
  & yes
  & yes \\
Long audio
  & no truncation
  & skip clips $>30$\,s \\
Language
  & \texttt{language L<asr\_text>}
  & decoder prefix + \texttt{transcribe} \\
\bottomrule
\end{tabular*}
\end{table}

Section~\ref{sec:targeted_selection} reports one-hour
adaptation of Qwen3-ASR-1.7B~\citep{shi2026qwen3asr}.
We repeat the same budgeted selection on Whisper-small~\citep{radford2023whisper}
to test whether the rule transfers to a weaker
encoder--decoder ASR model.
Difficulty scores are computed from Whisper-small itself,
so the selected subsets are not reused from the Qwen
experiment.
Because the Whisper encoder is limited to 30\,s, candidate
clips longer than 30\,s are skipped rather than truncated.
Truncating audio while retaining the full transcript raised
overall error from 36.84\% to 50--58\% in preliminary runs
and is not used here.
Selection otherwise matches the main experiment: a pool of
10{,}000 clips disjoint from the 1{,}000-recording
evaluation cohort, a $B=3600$\,s budget, and three
sampling seeds with the training RNG held fixed.

\subsection{Fine-tuning Settings}
We fully fine-tune each model for one epoch on the selected
one-hour subset, with no validation split during training.
Qwen3-ASR-1.7B follows the official Qwen3-ASR SFT recipe;
on 24\,GB GPUs we preserve its global batch of 256 by using
a per-device batch of 1 and 128 gradient-accumulation steps
in place of 32 and 4.
Whisper-small follows the public Hugging Face Whisper
fine-tuning recipe~\citep{gandhi2022whisperfinetuning}.
Table~\ref{tab:app_ft_hparams} lists the remaining settings.

\begin{table}[t]
\caption{Whisper-small recognition error after one-hour
fine-tuning.
Mean $\pm$ standard deviation over three sampling seeds.
$\Delta$ is the relative change versus the unadapted model.
Overall error pools English WER and Chinese/Thai CER by
reference units, as in Table~\ref{tab:targeted_selection}.}
\label{tab:app_whisper_selection}
\centering
\footnotesize
\setlength{\tabcolsep}{4pt}
\renewcommand{\arraystretch}{1.12}
\begin{tabular*}{\linewidth}{@{\extracolsep{\fill}}lccccc}
\toprule
Selection
  & seed 0
  & seed 1
  & seed 2
  & mean $\pm$ SD
  & $\Delta$ \\
\midrule
Original
  & 36.84 & -- & --
  & 36.84
  & -- \\
Random
  & 38.44 & 33.17 & 30.20
  & $33.94\pm4.18$
  & $-7.9$ \\
\midrule
Pool / SURE
  & 36.11 & 37.02 & 34.55
  & $35.90\pm1.25$
  & $-2.6$ \\
Pool / Captioner
  & 37.48 & 36.55 & 38.75
  & $37.59\pm1.10$
  & $+2.0$ \\
\midrule
Eval.\ / SURE
  & 33.44 & 30.43 & 31.08
  & $\mathbf{31.65\pm1.58}$
  & $\mathbf{-14.1}$ \\
Eval.\ / Captioner
  & 35.13 & 36.64 & 41.13
  & $37.63\pm3.12$
  & $+2.2$ \\
\bottomrule
\end{tabular*}
\end{table}

Eval.-guided SURE-Tagger is the only method that improves
on the unadapted model in all three seeds, reducing error
from 36.84\% to $(31.65\pm1.58)\%$, a 14.1\% relative
reduction.
Pool-guided SURE-Tagger yields only a small mean gain
($(35.90\pm1.25)\%$).
Random selection can also beat the baseline on average
($(33.94\pm4.18)\%$) but is unstable across seeds.
Both captioner-guided variants remain slightly worse than
the unadapted model.
The SURE-Tagger condition representation therefore
transfers beyond Qwen3-ASR-1.7B, while which of Eval.\ or
Pool is stronger can depend on the adapted model;
caption-derived attributes remain unhelpful under the same
budget.

\end{document}